\documentclass[journal]{IEEEtran}
\usepackage{cite}

\ifCLASSINFOpdf
\else
\usepackage[dvips]{graphicx}
\fi
\usepackage{url}

\usepackage{amssymb,amsmath,amsthm}
\usepackage{array}

\usepackage[small]{caption}
\usepackage{subcaption}
\usepackage[font=footnotesize]{subfig}

\usepackage{multirow}
\usepackage{psfrag}
\usepackage{color}
\usepackage[percent]{overpic}
\usepackage{booktabs}

\makeatletter
\def\th@plain{%
  \thm@notefont{}
  \itshape 
}
\def\th@definition{%
  \thm@notefont{}
  \normalfont 
}
\makeatother

\usepackage{xspace}
\newtheorem{theorem}{Theorem}
\newtheorem{lemma}{Lemma}

\newtheorem{proposition}{Proposition}
\newtheorem{corollary}{Corollary}
\theoremstyle{definition}
\newtheorem{definition}{Definition}
\theoremstyle{definition}
\newtheorem{remark}{Remark}
\DeclareMathOperator*{\argmax}{arg\,max}

\renewcommand{\P}{\mathbb{P}}
\newcommand{\R}{\mathbb{R}}
\newcommand{\E}{\mathbb{E}}
\newcommand{\ie}{\emph{i.e.,}\@\xspace}
\newcommand{\eg}{\emph{e.g.,}\@\xspace}
\newcommand{\deq}{\stackrel{\textnormal{d}}{=}}
\renewcommand{\L}{\mathcal{L}}
\renewcommand{\d}{\textnormal{d}}
\def\peq#1{\stackrel{\text{\scriptsize(#1)}}{=}}

\def\pleq#1{\stackrel{\text{\scriptsize(#1)}}{\leq}}

\graphicspath{{./Figs/}}

\begin{document}

\title{A Stochastic Geometry Analysis of Inter-cell Interference Coordination and Intra-cell Diversity}

\author{Xinchen~Zhang,~\IEEEmembership{Member,~IEEE,}
and~Martin~Haenggi,~\IEEEmembership{Fellow,~IEEE}  \vspace{-0.8cm}
\thanks{Manuscript date \today.

Xinchen Zhang ({\tt x.zhang@utexas.edu}) is with the Wireless Networking and Communications Group (WNCG),
University of Texas at Austin, TX, USA.

Martin Haenggi ({\tt mhaenggi@nd.edu}) is with the Department of Electrical Engineering,
University of Notre Dame, IN, USA.

This work was partially supported by the NSF (grants CNS 1016742
and CCF 1216407).

This work was presented in part at
2014 IEEE International Symposium on Information Theory (ISIT'14), Honolulu, HI, USA \cite{ZhangHaenggi2014ISIT}.
}}
\maketitle

\begin{abstract}

Inter-cell interference coordination (ICIC) and intra-cell diversity (ICD)
play important roles in improving cellular downlink coverage.
Modeling cellular base stations (BSs) as a homogeneous Poisson point process (PPP),
this paper provides explicit finite-integral expressions for the coverage probability
with ICIC and ICD, taking into account the temporal/spectral correlation
of the signal and interference.
In addition,
we show that in the high-reliability regime, where the user outage probability
goes to zero, ICIC and ICD affect the network coverage in drastically different ways:
ICD can provide \emph{order} gain while
ICIC only offers \emph{linear} gain.
In the high-spectral efficiency
regime where the SIR threshold goes to infinity,
the order difference in the coverage probability does not exist;
however a linear difference makes ICIC a better scheme than ICD
for realistic path loss exponents.
Consequently, depending on the SIR requirements, different combinations
of ICIC and ICD optimize the coverage probability.
\end{abstract}

\section{Introduction}

\subsection{Motivation and Main Contributions}

Recently, the Poisson point process (PPP) has been shown to be
a tractable and realistic model of cellular networks \cite{net:Andrews12tcom}.
However, the baseline PPP model predicts the coverage probability of the typical user to be less than 60\%
if the signal-to-interference-plus-noise ratio (SINR) is set to 0 dB---even if noise is neglected.
This is clearly insufficient to provide reasonable user experiences in the network.


To improve the user experiences, in cellular systems, 
the importance of inter-cell interference
coordination (ICIC) and intra-cell diversity (ICD) have long been recognized\cite{GoldsmithBook,tse2005fundamentals}.
Yet, so far, most of the PPP-based cellular analyses lack
a careful treatment of these two important aspects of the network,
partly due to the lack of a well-established approach to deal with
the resulting temporal or spectral correlation \cite{HaenggiSpaSWiN13}.


Modeling the cellular network as a homogeneous PPP,
this paper explicitly takes into account the temporal/spectral correlation
and analyzes the benefits of ICIC and ICD in cellular downlink under idealized assumptions.
Consider the case where a user is always
served by the BS that provides the strongest signal averaged over small-scale fading but
not shadowing\footnote{Without shadowing, this is the nearest BS association policy
as used, for example, in \cite{net:Andrews12tcom}.}.
For ICD, we consider the case where the serving BS always transmit to the user in $M$ resource blocks (RBs)
simultaneously and the user always decodes from the RB with the best SIR (selection combining).
For ICIC, we assume under $K$-BS coordination, the RBs that the user is assigned
are silenced at the next $K-1$ strongest BSs.

Note that
both of the schemes create extra load (reserved RBs) in the network:
ICIC at the adjacent cells and ICD at the serving cell.
Therefore, it is important to quantify the benefits of
ICIC and ICD in order to design efficient systems.
The main contribution of this paper is to provide explicit expressions
for the coverage probability with $K$-BS coordination and $M$-RB selection combining.
Notably, we show that,
in the high-reliability regime, where the outage probability goes to zero,
the coverage gains due to ICIC and ICD are qualitatively different:
ICD provides \emph{order} gain while ICIC only offers \emph{linear} gain.
In contrast, in the high-spectral efficiency
regime, where the SIR threshold goes to infinity,
such order difference does not exist 
and ICIC usually offers larger (linear) gain than ICD in terms of coverage probability.
The techniques presented in this paper have the potential to
lead to a better understanding of the performance of more complex
cooperation schemes in wireless networks, which inevitably involve
temporal or spectral correlation.

\subsection{ICIC, ICD and Related Works}

Generally speaking, inter-cell interference coordination (ICIC)
assigns different time/frequency/spatial dimensions to users from
different cells and thus reduces the inter-cell interference.
Conventional ICIC schemes are mostly based on the idea of frequency reuse.
The resource allocation under cell-centric ICIC is designed offline and does not depend on the
user deployment.
While such schemes are advantageous due to their simplicity and small
signaling overhead, they are clearly suboptimal since the pre-designed frequency reuse pattern cannot
cope well with the dynamics of user distribution and channel variation.
Therefore, there have been significant efforts in facilitating ICIC schemes,
where the interference coordination (channel assignment) is based on real user locations
and channel conditions and enabled by multi-cell coordination.
Different user-centric (coordination-based) ICIC schemes in OFDMA-based networks are well summarized in
the recent survey papers \cite{FodorKRRSM09,HamzaElsayed13,KostaTafazolliICST2013}.

Conventionally, most of the performance analyses of ICIC are based on network-level
simulation, and the hexagonal-grid model is frequently used \cite{HamzaElsayed13}.
Since real cellular deployments are subject to many practical constraints,
recently more and more analyses are based on randomly distributed BSs,
mostly using the PPP as the model.
These stochastic geometry-based models not only provide alternatives to the classic grid models
but also come with extra mathematical tractability \cite{net:Andrews12tcom,net:ElSawy13tut,net:mh12}.
In terms of the treatment of ICIC, the most relevant papers to this one are
\cite{GantiTCOMM2012,ganti:novlan:transwireless:2011:march,MadhusudhananBrown2012Globecom},
where the authors analyzed partial frequency reuse schemes using independent thinning.
The authors in \cite{KHuangAndrewsIT2013,AkoumHeath2012SPAWC} considered BS coordination
based on clusters grouped by tessellations. 
Different from these papers, this paper focuses on user-centric ICIC schemes
where the spatial correlation of the coordinated cells is explicitly accounted for.

It is worth noting that ICIC is closely related to multi-cell processing (MCP) and coordinated
multipoint (CoMP) transmission,
see \cite{GesbertYuJSAC2010,JZhangAndrewsTWC2008,KHuangAndrewsIT2013,AkoumHeath2012SPAWC} and the references therein.
MCP/CoMP emphasizes the multi-antenna aspects of the cell coordination, 
while the form of ICIC considered in this paper does not take into account
the use of MIMO (joint transmission) techniques and thus is not subject to the
considerable signaling and processing overheads of
typical MCP/CoMP schemes, which include
symbol-level synchronization and joint precoder design \cite{KostaTafazolliICST2013}.
Thus it can be considered as a simple form of MCP/CoMP that is light on overhead.

Intra-cell diversity (ICD) describes the diversity gain achieved by having
the serving BS opportunistically assigns users with their best channels.
In cellular systems, diversity exists in space,
time, frequency and among users\cite{tse2005fundamentals}.
It is well acknowledged that diversity can significantly boost the network coverage.
However, conventional analyses of diversity usually do not include
the treatment of interference, \eg \cite{net:Viswanath02,SongLi2005TWC}.

In order to analytically characterize diversity in wireless networks with interference,
a careful treatment of interference correlation is necessary, 
otherwise the results may be misleading.
Therefore, there have been a few recent efforts in understanding this correlation\cite{net:Haenggi14twc,net:Haenggi12cl,SchilcherBrandnerTMC2012,TanbourgiDhillonAndrewsJondral2014,ganti:globecom3:2012,net:Gong12tmc}.
Notably, \cite{net:Haenggi14twc} shows that in an \emph{ad hoc} type network,
simple retransmission schemes do not result in diversity gain if interference correlation is considered\footnote{
Different from conventional SNR-based diversity analysis,
\cite{net:Haenggi14twc} calculates the diversity gain by considering the case
where signal to interference ratio (SIR) goes to infinity, which is an analog of the classic (interference-less) notional of diversity.
This paper follows the same analogy.}.
Analyzing the intra-cell diversity (ICD) under interference correlation,
this paper shows that a diversity gain \emph{can} be obtained
in a cellular setting where the receiver is always connected to the strongest BS,
in sharp contrast with the conclusion drawn from \emph{ad hoc} type networks in \cite{net:Haenggi14twc}.


\subsection{Paper Organization}

	%
%

The rest of the paper is organized as follows:
Section~\ref{sec:sysPLPSPF} presents the system model
and discusses the comparability of ICIC and ICD.
Sections~\ref{sec:ICIC}~and~\ref{sec:ICD} derive the coverage probability
for the case with ICIC or ICD only, respectively, and provide results on the asymptotic
behavior of the coverage probability in the high-reliability
as well as high-spectral efficiency regimes.
The case with both ICIC and ICD is analyzed in Section~\ref{sec:ICIC&ICD}.
We validate our model and discuss fundamental trade-offs between ICIC and ICD in Section~\ref{sec:numer}.
The paper is concluded in Section~\ref{sec:conclu}.

\section{System Model, the Path Loss Process with Shadowing (PLPS) and the Coverage Probability \label{sec:sysPLPSPF}}

\subsection{System Model}

Considering the typical user at the origin $o$,
we use a homogeneous Poisson point process (PPP) $\Phi\subset \R^2$
with intensity $\lambda$ to model
the locations of BSs on the plane.
To each element of the ground process $x\in \Phi$,
we add independent marks\footnote{For analytical tractability, the \emph{spatial} shadowing correlation due to common obstacles
is not considered in this model.} $S_x\in \R^+$ and $h^m_x\in \R^+$, where $m\in [M]$ and $M\in\mathbb{N}$,\footnote{We use
$[n]$, to denote the set $\{1,2,\cdots,n\}$.} to denote
the (large-scale) shadowing and (power) fading effect on the link from $x$ to $o$ at the $m$-th resource block (RB),
and the combined (marked) PPP is denoted as
$\hat\Phi = \{(x_i, S_{x_i}, (h^m_{x_i})_{m=1}^M)\}$.
In particular, under power law path loss,
the received power at the typical user $o$ at the $m$-th RB from a BS at $x\in\Phi$
is
\begin{equation}
	P_x = S_x h^m_x \|x\|^{-\alpha},
\end{equation}
where $\alpha$ is the path loss exponent.
In this paper, we focus on Rayleigh fading, \ie $h_x$ is exponentially
distributed with unit mean but allow the shadowing distribution to
be (almost) arbitrary.

\begin{figure}[t]
\centering
\includegraphics[width=7cm]{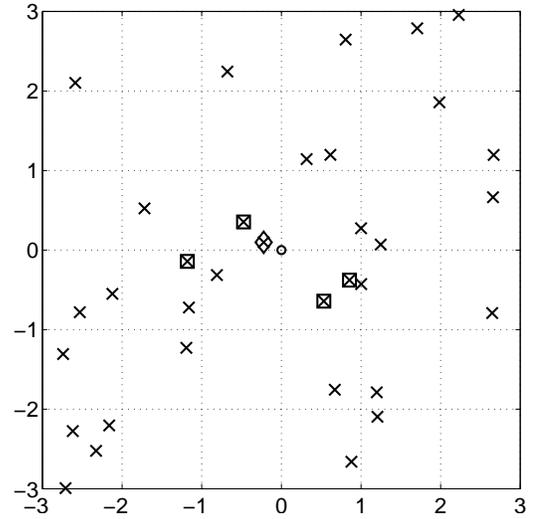}
\caption{A realization of the cellular network modeled by a homogeneous PPP $\Phi$.
The network is under $K$-BS ($K=5$) coordination with lognormal shadowing. 
The typical user is denoted by $\circ$,
the BSs by $\times$, the serving BS by $\Diamond$ and the coordinated
non-serving BS by $\Box$.
}
\label{fig:ICICIll}
\end{figure}

Fig.~\ref{fig:ICICIll} shows a realization of a PPP-modeled
cellular network under $K$-BS coordination with lognormal shadowing.
Due to the shadowing effect, the $K$ strongest BSs under coordination are
not necessarily the $K$ nearest BSs.

The base station locations (ground process $\Phi$) and the shadowing
random variables $S_x$ are static over time and frequency (i.e., over all RBs),
which is the main reason of the spectral/temporal correlation of signal and interference.
In comparison, the (small-scale) fading $h_x^m$ is iid over RBs.
Both $S_x$ and $h_x^m$ are iid over space (over $x$).


The user is assumed to be associated with the strongest (without fading) BS
and is called \emph{covered} (without ICIC) at the $m$-th RB iff
\begin{equation}
	\textnormal{SIR}_m = \frac{S_{x_0} h^m_{x_0} \|x_0\|^{-\alpha}}{\sum_{y\in\Phi\setminus \{x_0\}} S_y h^m_y \|y\|^{-\alpha}} > \theta,
	\label{equ:cvrg0}
\end{equation}
where $x_0 = \argmax_{x\in\Phi} S_x \|x\|^{-\alpha}$ is the serving BS.

\subsection{The Path Loss Process with Shadowing (PLPS)}

\begin{definition}[The path loss process with shadowing]
The path loss process with shadowing (PLPS) $\Xi$ is the point process on $\R^+$
mapped from $\hat\Phi$, where $\Xi = \{\xi_i = \frac{\|x\|^\alpha}{S_x}, x\in\Phi\}$
and the indices $i\in\mathbb{N}$ are introduced such that $\xi_k<\xi_j$ for all $k<j$.
\end{definition}

Note that the PLPS is an \emph{ordered} process.
It captures the effect of shadowing and spatial node distribution
of the network at the same time, and consequently, determines the BS association.

\begin{lemma}
	The PLPS $\Xi$ is a one-dimensional PPP with intensity measure
	$\Lambda((0,r])=\lambda \pi r^\delta \E[S^\delta]$, where $\delta=2/\alpha$,
	$S\deq S_x$ and $\deq$ means equality in distribution.
\label{lem:PLPSisPPP}
\end{lemma}

The proof of Lemma~\ref{lem:PLPSisPPP} is analogous to 
that of \cite[Lemma 1]{Zhang12globecom}
and is omitted from the paper.
The intensity measure of the PLPS demonstrates the necessity of the $\delta$-th moment
constraint on the shadowing random variable $S_x$.
Without this constraint, the aggregate received power (with or without fading)
is unbounded almost surely.

\subsection{The Coverage Probability and Effective Load Model}

Similar to the construction of $\hat\Phi$,
We construct a marked PLPS $\hat\Xi = \{(\xi_i, (h^m_{\xi_i})_{m=1}^M,\chi_{\xi_i})\}$,
where we put two marks on each element of the PLPS $\Xi$:
$h^m_\xi = h^m_x,\; m\in[M],\; x\in\Phi$, are the iid fading random variables directly mapped from
$\hat\Phi$;
$\chi_{\xi}\in\{0,1\}$ indicates whether a BS represented by $\xi$
is transmitting at the RB(s) assigned to the typical user\footnote{
It is assumed that the RBs are grouped into chunks of size $M$,
\ie each BS either transmits at all the $M$ RBs or does not transmit at any of these RBs.}.
In the case where no ambiguity is introduced, we will use $h^m_i$ as an abbreviation for $h^m_{\xi_i}$
and $\chi_i$ as a short of $\chi_{\xi_i}$.
For example, if no ICIC is considered, 
we have $\chi_i=1,\;\forall i$,\footnote{We assume all the BSs are fully loaded,
\ie each RB is either used in downlink transmission or silenced due to coordination.}
and the coverage condition in \eqref{equ:cvrg0} can be written in terms of the marked PLPS
as
\begin{equation}
	\textnormal{SIR}_m = \frac{h^m_1 \xi_1^{-1}}{\sum_{i=2}^\infty h^m_i \xi_i^{-1}} >\theta.
\end{equation}

With ICIC, the value of $\chi_i$ is determined by the scheduling policy.
Given $\chi_i$, the coverage condition at the $m$-th RB under $K$-BS coordination
can be expressed in terms of the marked PLPS as
\begin{equation}
	\textnormal{SIR}_{K,m} = \frac{h^m_1 \xi_1^{-1}}{\sum_{i=2}^\infty \chi_i h^m_i \xi_i^{-1}} >\theta.
\end{equation}
By $K$-BS coordination (ICIC), we assume the $K-1$ strongest non-serving BSs of the typical user
do not transmit at the RBs to which the user is assigned\footnote{This can be implemented by letting the UE to identify
the $K$ strongest BSs and then reserve the RBs at all of them.}.
Thus, we have $\chi_i = 0,\;\forall i\in [K]\setminus \{1\}$.\footnote{By default $\chi_1 = 1$.}
For $i>K$, the exact value of $\chi_i$ is hard to model
since the BSs can either transmit to its own users in the RB(s) assigned to the typical
user or reserve these RB(s) for users in nearby cells,
and the muted BSs can effectively ``coordinate" with multiple serving BSs at the same time.
Therefore, the resulting density of the active BSs outside the $K$ coordinating BSs
is a complex function of the user distribution, (joint) scheduling algorithms
and shadowing distribution.

In order to maintain tractability, we assume $\chi_i, \; i>K$ are iid Bernoulli random variables with
(transmitting) probability $1/\kappa$.
Such modeling is justified by the random distribution of the users and the shadowing
effect \cite{Bar13Lattice}.
Here, $\kappa\in[1,K]$ is called the \emph{effective load} of ICIC.
$\kappa = K$ implies all the coordinating BS clusters do not overlap
while $\kappa = 1$ represents the scenario where all the users assigned to the same RB(s)
in the network share the same $K-1$ muted BSs.
The actual value of $\kappa$ lies between these two extremes\footnote{
The statement is true under the full-load assumption.
In the case where some cells may contain no users, it is possible that $\kappa>1$ while $K=1$.
But this does not have a large influence on the accuracy of the analyses as is shown
in detail in Section~\ref{sec:numer}.}
and is determined by the scheduling procedure which this paper does not explicitly study.
However, we assume that $\kappa$ is known. 
The accuracy of this model will be validated in Section~\ref{sec:numer}.


Let ${\bf S}_{K,m}\triangleq \{\textnormal{SIR}_{K,m}>\theta\}$ be the event of coverage at the $m$-th RB.
We consider the coverage probability with inter-cell interference coordination (ICIC)
and intra-cell diversity (ICD) formally defined as follows.

\begin{definition}
The coverage probability with $K$-BS coordination and $M$-RB selection combining
is
\begin{equation*}
\mathsf{P}^{\textnormal{c}}_{K,M} = \mathsf{P}^{\cup \textnormal{c}}_{K,M} \triangleq \P(\cup_{m=1}^M {\bf S}_{K,m}).
\end{equation*}
\end{definition}

In other words, the typical user is covered iff the received SIR at any of the $M$ RBs is greater than $\theta$.
The superscript c denotes \emph{coverage} and $\cup$ stresses that $\mathsf{P}^{\cup \textnormal{c}}_{K,M}$
is the probability of being covered in \emph{at least one of the M RBs}.
(If there is no possibility of confusion, we will use $\mathsf{P}^{\cup \textnormal{c}}_{K,M}$
and $\mathsf{P}^{\textnormal{c}}_{K,M}$ interchangeably.)

\subsection{System Load and Comparability}

In the baseline case without ICIC and ICD, 
each user occupies a single RB at the serving BS.
With (only) $M$-RB selection combining, each user occupies $M$ RBs at the serving BS.
Thus, the system load is increased by a factor of $M$.
The load effect of ICIC can be described by the effective load $\kappa$
since, as discussed above, in a network with $K$-BS coordination there are
$1/\kappa$ of the BSs actively serving the users in a single RB whereas
each BS serves one user in every RB in the baseline case,
\ie the load is increased by a factor of $\kappa$ due to ICIC.
The fundamental comparability of ICIC and ICD comes from the similarity
in introducing extra load in the system and will be explored in more detail in Section~\ref{sec:numer}.

\section{Intercell Interference Coordination (ICIC)\label{sec:ICIC}}

This section focuses on the effect of ICIC
on the coverage probability.
Since no ICD is considered,
we will omit the superscript $m$ on the fading random variable $h^m_\xi,\;\xi\in\Xi$, for simplicity.

\subsection{Integral Form of Coverage Probability}

Our analysis will be relying on a statistical property of the marked
PLPS $\hat\Xi$ stated in the following lemma.

\begin{lemma}
	For $\hat\Xi = \{(\xi_i, h_i, \chi_i)\}$,
	let $X_k = \xi_1/\xi_k$
	and $Y_k = \xi_k^{-1}/I_k$, where $I_k \triangleq \sum_{i=k+1}^{\infty} \chi_i h_i \xi_i^{-1}$.
	For all $k\in\mathbb{N}$, the two random variables $X_k$ and $Y_k$ are independent.
\label{lem:2worlds}
\end{lemma}

\begin{IEEEproof}
	If $k=1$, the lemma is trivially true, since $X_1\equiv 1$ 
	while $Y_1$ has some non-degenerate distribution.
	
	For $k\geq 2$, $x_1\in[0,1]$ and $x_2\in\R^+$, the joint ccdf of $\xi_1/\xi_k$ and $\xi_k/I_k$ can be expressed as
	\begin{align*}
		\hspace{2em}&\hspace{-2em}\P(X_k > x_1, Y_1 > x_2)	\\
			&= \E_{\xi_k}\left[\P\left(\frac{\xi_1}{\xi_k} > x_1, \frac{\xi_k^{-1}}{I_k} > x_2\right) \mid \xi_k\right]		\\
			&\peq{a} \E_{\xi_k}\left[\P\left(\frac{\xi_1}{\xi_k }> x_1\right)\P\left(\frac{\xi_k^{-1}}{I_k} > x_2\right) \mid \xi_k\right]	\\
			&\peq{b} \P\left(\frac{\xi_1}{\xi_k} > x_1\right) \E_{\xi_k}\left[\P\left(\frac{\xi_k^{-1}}{I_k} > x_2\right) \mid \xi_k\right]	\\
			&= \P\left(\frac{\xi_1}{\xi_k} > x_1\right)\P\left(\frac{\xi_k^{-1}}{I_k} > x_2\right),
	\end{align*}
	where (a) is due to the fact that
	$\{\xi_i, i < k\}$ and $\{\xi_i, i > k\}$ are conditionally independent given $\xi_k$ by the Poisson property
	and $\{h_i\}$, $\{\chi_i\}$ are iid and independent from $\Xi$.
	(b) holds since, conditioning on $\xi_k$ implies that there are $k-1$ points on $[0,\xi_k)$.
	Thus, thanks to the Poisson property, it can be shown that given $\xi_k$, $\xi_1/\xi_k$ follows the same distribution
	as that of the minimum of $k-1$ iid random variables with cdf $\min\{x^\delta,1\}\mathsf{1}_{\R^+}(x)$.\footnote{
	In fact, for general inhomogeneous PPP on $\R^+$ of intensity measure $\Lambda(\cdot)$,
	given there are $N$ points on $[0,x_0)$ the joint distribution of the locations of the $N$ points
	is the same as that of $N$ iid random variables with cdf $\Lambda([0,x))/\Lambda([0,x_0))$ \cite[Theorem 2.25]{net:mh12}.}
	Since the resulting conditional distribution of $\xi_1/\xi_k$ does not depend on $\xi_k$,
	this distribution is also the marginal distribution of $\xi_1/\xi_k$ as is stated in the lemma.
\end{IEEEproof}

Furthermore, due to Lemma~\ref{lem:PLPSisPPP},
it is straightforward to obtain the ccdf of $\xi_1/\xi_k,\;\forall k\geq 2$,
which is formalized in the following lemma.

\begin{lemma}
For all $k\in\mathbb{N}\setminus\{1\}$, The ccdf of $\xi_1/\xi_k$ is
\begin{equation*}
	\P\left(\frac{\xi_1}{\xi_k} > x\right) = (1-x^{\delta})^{k-1}, \; x\in[0, 1].
\end{equation*}
\label{lem:xi1xik}
\end{lemma}
\begin{IEEEproof}
As discussed in the proof of Lemma~\ref{lem:2worlds},
by the Poisson property and the intensity measure of $\Xi$ given in
Lemma~\ref{lem:PLPSisPPP},
conditioned on $\xi_k, k > 1$,
$\xi_i/\xi_k\deq X_{i:k-1}$, where $X_{i:k-1}$
denotes the $i$-th order statistics of $k-1$ iid
random variables with cdf $\P(X<x) = \min\{x^\delta,1\}\mathsf{1}_{\R^+}(x)$.
Thus, $\P(\frac{\xi_1}{\xi_k} > x)$ is the probability
that all the $k-1$ iid random variables are larger than $x$.
\end{IEEEproof}

\begin{lemma}
For $\hat\Xi = \{(\xi_i, h_i)\}$, 
let
\[I_\rho = \sum_{\xi\in\Xi\cap(\rho,\infty)} \chi_\xi h_{\xi} \xi^{-1}\]
for $\rho>0$.
The Laplace transform of $\rho I_\rho$ is
\begin{equation}
	\L_{\rho I_\rho} (s) = \exp\left( -\frac{\lambda}{\kappa}\pi\E[S^\delta] C(s) \rho^\delta \right),
\end{equation}
where $C(s) = \frac{s\delta}{1-\delta} {_2 F_1} (1,1-\delta;2-\delta;-s) $ and $_2 F_1 (a,b;c;z)$ is the
Gauss hypergeometric function.
\label{lem:LtildeIrho}
\end{lemma}

\begin{IEEEproof}
First, we can calculate the Laplace transform of $I_\rho$ using the probability generating
functional (PGFL) of PPP \cite{net:mh12}, \ie
\begin{align*}
	\L_{I_\rho} (s) &= \E[\exp(-s\sum_{\xi\in\Xi\cap(\rho,\infty)} \chi_\xi h_\xi \xi^{-1})]		\\
									&=	\E_{\Xi}\prod_{\xi\in \Xi \cap(\rho,\infty)} \E_{h,\chi}[\exp(-s\chi h\xi^{-1})]  \\
									&= \exp\left( - \E_{h,\chi} \left[\int_\rho^\infty (1-e^{-s\chi h/x}) \Lambda(\d x)\right] \right),
\end{align*}
where $\chi$ is a Bernoulli random variable with mean $\frac{1}{\kappa}$,
$\Lambda(\cdot)$ is the intensity measure of $\Xi$ and by Lemma~\ref{lem:PLPSisPPP},
$\Lambda(\d x)=\lambda\pi\E[S^\delta]\delta x^{\delta-1} \d x$.
Then, straightforward algebraic manipulation yields
\begin{multline*}
	\L_{I_\rho} (s) =\\  \exp\left( - \frac{\lambda}{\kappa} \pi\E[S^\delta]\E_h\left[ (sh)^\delta \gamma(1-\delta,\frac{sh}{\rho})
	- \rho^\delta(1-e^{-\frac{sh}{\rho}}) \right] \right).
\end{multline*}
Since, for an arbitrary random variable $X$ and constant $u$,
$\L_{uX}(s) \equiv \L_X(us)$, we have
\begin{equation}
	\L_{\rho I_\rho} (s) = \L_{I_\rho} ({s\rho}) = \exp\left( -\frac{\lambda}{\kappa}\pi\E[S^\delta] C(s) \rho^\delta \right),
\end{equation}
where $C(s) = \E_h[(sh)^\delta\gamma(1-\delta,sh)+e^{-sh}-1]$.
The proof is completed by considering the exponential distribution of $h$.
\end{IEEEproof}

Here,
$I_\rho$ can be understood as the interference from
BSs having a (non-fading) received power weaker than $\rho^{-1}$.
In the case without shadowing, \ie $S_x\equiv 1$, it can also be
understood as the interference coming from outside a disk centered
at the typical user with radius $\rho^{\frac{1}{\alpha}}$.

\begin{lemma}
The Laplace transform of $\xi_k I_k$ is
\begin{equation}
	\L_{\xi_k I_k}(s) = \frac{1}{\left(C_\kappa(s,1)\right)^k},
	\label{equ:xikTildIkLaplace}
\end{equation}
where $C_\kappa(s,m) = \frac{\kappa-1}{\kappa}+\frac{1}{\kappa} {_2F_1}(m,-\delta;1-\delta;-s)$.
\label{lem:xikTildIkLaplace}
\end{lemma}

\begin{IEEEproof}
First, the pdf of $\xi_k$ can be derived
analogously to the derivation of \cite[Lemma 3]{Zhang12globecom}
as
\begin{equation*}
	f_{\xi_k}(x) = (\lambda\pi\E[S^\delta])^k\frac{\delta x^{k\delta-1}}{\Gamma(k)}\exp\left(-\lambda\pi\E[S^\delta] x^\delta\right).
\end{equation*}

Then, thanks to Lemma~\ref{lem:LtildeIrho}, 
the Laplace transform of $\xi_k I_{\xi_k}$ can be obtained by
deconditioning $\L_{\rho I_\rho}(s)$ (given $\rho$)
over the distribution of $\xi_k$. This leads to
\begin{equation*}
	\L_{\xi_k I_k}(s) = \frac{1}{\left(1+\frac{1}{\kappa}C(s)\right)^k},
\end{equation*}
where $1+\frac{1}{\kappa} C(s) = C_\kappa(s,1)$. 
\end{IEEEproof}

Note that
although the path loss exponent $\alpha$ is not explicitly taken as a parameter of $C_\kappa(\cdot,\cdot)$,
$C_\kappa(\cdot,\cdot)$ depends on $\alpha$ by definition.
Thus, the value of $\alpha$ affects all the results.

Since we consider Rayleigh fading,
the coverage probability without ICIC is just the Laplace
transform of $\xi_1 I_1$.
The special case of $k=\kappa=1$ of Lemma~\ref{lem:xikTildIkLaplace} 
corresponds to the well-known coverage probability
in cellular networks (under the PPP model) without ICIC or ICD  \cite{net:Andrews12tcom}, 
\begin{equation*}
	\mathsf{P}^\textnormal{c}_{1,1} = \frac{1}{C_1(\theta,1)}.	
\end{equation*}
%
%
Note that since we consider the full load case and $\kappa\in[1,K]$,
$K=1$ implies $\kappa=1$.
In the more general case $K>1$,
$\kappa$ depends on the user distribution and the scheduling policy
and thus is hard to determine.
However, treating $\kappa$ as a parameter we obtain the following theorem
addressing the case with non-trivial coordination.

\begin{theorem}[$K$-BS coordination]
	The coverage probability for the typical user under $K$-cell coordination ($K>1$) is
	\begin{equation}
		\mathsf{P}^\textnormal{c}_{K,1} = (K-1) \int_0^1 \frac{(1-x^\delta)^{K-2} \delta x^{\delta-1}}{\left(C_\kappa(\theta x, 1)\right)^{K}} \d x,
	\label{equ:Kcoor-integral}
	\end{equation}
where $C_\kappa(s,m) = \frac{\kappa-1}{\kappa}+\frac{1}{\kappa} {_2F_1}(m,-\delta;1-\delta;-s)$.\label{thm:KCoord}
\end{theorem}

\begin{IEEEproof}
The coverage probability can be written in terms of the PLPS as
\begin{align}
	\mathsf{P}^\textnormal{c}_{K,1} = \P(h_1 \xi_1^{-1} > \theta I_K) = \P\left(\frac{h_1 \xi_K^{-1}}{I_K} > \theta \frac{\xi_1}{\xi_K}\right),
\end{align}
where $h_1$ is exponentially distributed with mean 1, and thus
$\P(\frac{h_1 \xi_K^{-1}}{I_K} > x) = \L_{\xi_K I_K}(x)$.
Since ${h_1 \xi_K^{-1}}/{I_K}$ and ${\xi_1}/{\xi_K}$ are statistically independent (Lemma~\ref{lem:2worlds}),
we can calculate the coverage probability by
\begin{equation}
	\mathsf{P}^\textnormal{c}_{K,1} = \int_0^1 \L_{\xi_K I_K}(\theta x) \d F_{\xi_1/\xi_K}( x),
\end{equation}
where $F_{\xi_1/\xi_K} (x) = 1-(1-x^{\delta})^{K-1}$ is the cdf of $\xi_1/\xi_K$
given by Lemma~\ref{lem:xi1xik}.
The theorem is thus proved by change of variables.
\end{IEEEproof}

The finite integral in (\ref{equ:Kcoor-integral}) can be straightforwardly
evaluated numerically.

\begin{remark}
The Gauss hypergeometric function can be cumbersome to evaluate numerically,
especially when embedded in an integral, as in \eqref{equ:Kcoor-integral}.
Alternatively, $C_1$ can be expressed as 
\begin{equation*}
	C_1(s,m)=\frac{1}{(s+1)^m}+s^\delta m\textnormal{B}^\textnormal{u}_{\frac{1}{s+1}}(m+\delta,1-\delta),
\end{equation*}
and $C_\kappa(s,m) = \frac{\kappa-1}{\kappa}+\frac{1}{\kappa}C_1(s,m)$. Here,
 $\textnormal{B}^\textnormal{u}_x(a,b) = \int_x^{1} y^{a-1} (1-y)^{b-1} \d y$, $x\in[0,1]$,
is the \emph{upper} incomplete beta function, which can be calculated much more efficiently
in many cases. In Matlab, the speed-up compared with the hypergeometric function is at least
a factor of 30.

\end{remark}

\begin{figure}[t]
\centering
\psfrag{Pc}[c][t]{$\mathsf{P}^\textnormal{c}_{K,1}$}
\psfrag{kapeqK}[c][c]{$\kappa=K$}
\psfrag{kapeq1}[c][c]{$\kappa= 1$}
\begin{overpic}[width=\linewidth]{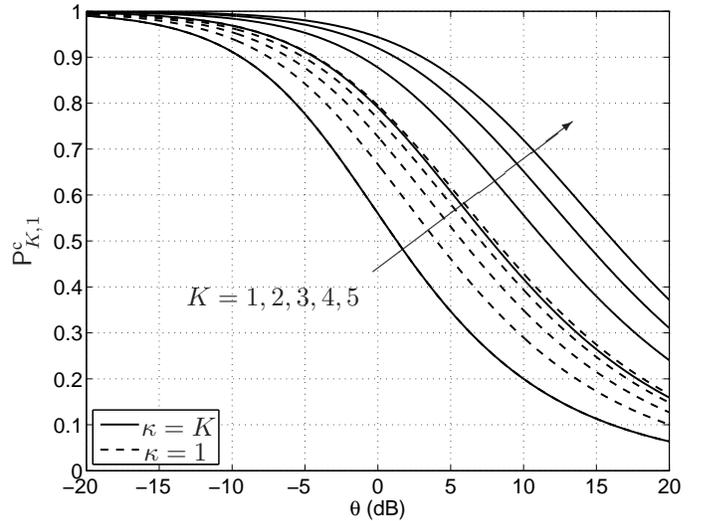}
	\put(53,38){\vector(4,3){30}}
	\put(25,33){$K=1,2,3,4,5$}
\end{overpic}
\caption{The coverage probability under $K$-BS coordination, $\mathsf{P}^\textnormal{c}_{K,1}$, for $K=1,2,3,4,5$
(lower to upper) and $\kappa=K$ and $\kappa= 1$. The path loss exponent $\alpha = 4$.
When $K=1$, the dashed line and solid line overlap.}
\label{fig:KcoordCoverage}
\end{figure}

Fig.~\ref{fig:KcoordCoverage} demonstrates the effect of ICIC
on the coverage probability for $\kappa= 1$ and $\kappa = K$.
The former case may be interpreted as a lower bound
and the latter case an upper bound.
As expected, the larger $K$,
the higher the coverage probability for all $\theta$.
On the other hand, the marginal gain of cell coordination
decreases with increasing $K$ since the interference, if any, from
far away BSs is attenuated by the long link distance
and affects the SIR less.

Fig.~\ref{fig:KcoordCoverage} also shows that larger $\kappa$ results
in larger coordination gain in terms of SIR.
This is due to the fact that coordination not only mutes the strongest $K-1$
interferers but also thins the interfering BSs outside the coordinating cluster.
However, this does not mean that the system will be better-off by implementing a larger $\kappa$.
Instead, from the load perspective,
the (SIR) gain is accompanied with the loss in bandwidth (increased load) since fewer
BSs are actively serving users.
The SIR-load trade-off will be further discussed along with the model validation in Section~\ref{sec:numer}.


\subsection{ICIC in the High-Reliability Regime}

While the finite integral expression given in Theorem~\ref{thm:KCoord}
is easy to evaluate numerically, it is also desirable to find a simpler
estimate that lends itself to a more direct interpretation of the benefit of ICIC.
This subsection investigates the asymptotic behavior of ICIC when $\theta \to 0$.
Note that $\theta\to 0$ refers to the high-reliability regime since
in this limit the typical user is covered almost surely.

In practice, the high-reliability regime ($\theta\to 0$)
is usually where the control channels operate.
In the LTE system (narrowband), the lowest MCS mode for downlink transmission supports an SINR about -7~dB
and thus may also be suitable for the high-reliability analysis.
In wide-band systems (\eg CDMA, UWB),
the system is more robust against interference and noise. Thus, $\theta$ is much smaller
and the high-reliability analysis is more applicable.

\begin{proposition}
Let $\mathsf{P}^\textnormal{o}_{K,1} = 1-\mathsf{P}^\textnormal{c}_{K,1}$
be the outage probability of the typical user for $K\in\mathbb{N}$. Then,
\begin{equation}
	\mathsf{P}^\textnormal{o}_{K,1} \sim a_K \theta, \text{ as } \theta\to 0,
	\label{equ:PoK1Sim}
\end{equation}
where  \[a_K = \frac{1}{\kappa}\frac{K!}{(1+\delta^{-1})_{K-1}}\frac{\delta}{1-\delta} \]
and $(x)_n = \prod_{i=0}^{n-1} (x+i)$ is the (Pochhammer) rising factorial.
\label{prop:PoK1theta->0}
\end{proposition}
\begin{IEEEproof}
See Appendix \ref{app:PoK1theta->0}.
\end{IEEEproof}

\begin{figure}[t]
\centering
\psfrag{ak}[c][t]{$a_K$}
\begin{overpic}[width=\linewidth]{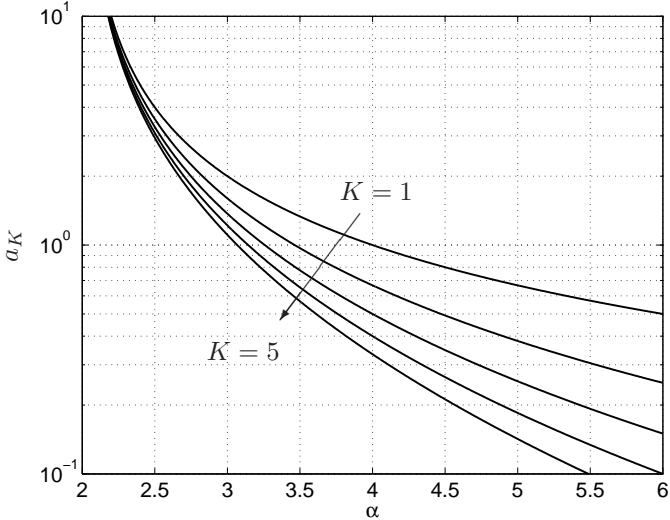}
	\put(30,25){$K=5$}
	\put(50,49){$K=1$}
	\put(53,47){\vector(-3,-4){12}}
\end{overpic}
\caption{The asymptotic coverage probability coefficient $a_K$ from Proposition~\ref{prop:PoK1theta->0}
as a function of the path loss exponent $\alpha$ under $K$-cell coordination (for $K=1,2,3,4,5$, upper to lower).}
\label{fig:aKalpha}
\end{figure}

Proposition~\ref{prop:PoK1theta->0}
shows that for pure ICIC schemes, the number of coordinating BSs only linearly affects the outage probability
in the high-reliability regime.
However, depending on the value of $\theta$,
even the linear effect may be significant.
In Fig.~\ref{fig:aKalpha}, we plot the coefficient $a_K$ for $K = 1,2,3,4,5$
as a function of the path loss exponent $\alpha$,
assuming $\kappa= 1$.
The difference (in ratio) between $a_K$ for different $K$
indicates the usefulness of ICIC, and this figure shows that
ICIC is more useful when the path loss exponent $\alpha$ is large.
This is consistent with intuition, since the smaller the path loss exponent,
the more the interference depends on the far-away interferers
and thus the less useful the local interference coordination is.
For other values of $\kappa$, the same trend can be observed.

\subsection{ICIC in High Spectral Efficiency Regime}

The other asymptotic regime is when $\theta \to \infty$.
In this regime, the coverage probability goes to zero while the spectral
efficiency goes to infinity.
Thus, it is of interest to study how the coverage probability decays with $\theta$.

\begin{proposition}
The coverage probability of the typical user for $K$-BS ($K>1$) coordination
satisfies 
\begin{equation}
	\mathsf{P}^\textnormal{c}_{K,1} \sim b_K \theta^{-\delta}, \text{ as } \theta\to \infty,
	\label{equ:PcK1SimHSE}
\end{equation}
where $b_K = (K-1)\int_0^\infty\frac{\delta x^{\delta-1}}{\left(C_\kappa(x,1)\right)^K}\d x$.
\label{prop:PcK1theta->Inf}
\end{proposition}

\begin{IEEEproof}
We prove the proposition by studying the asymptotic behavior of $\theta^\delta \mathsf{P}^\textnormal{c}_{K,1}$.
Using Theorem~\ref{thm:KCoord} and a change of variable,
we have 
\begin{equation}
	\theta^\delta \mathsf{P}^\textnormal{c}_{K,1}
	= (K-1) \int_0^\theta \frac{\delta x^{\delta-1}(1- x^\delta/\theta^\delta)^{K-2}}{\left(C_\kappa(x,1)\right)^K} \d x.
\end{equation}
Considering the sequence of functions (indexed by $\theta$)
\begin{equation*}
	f_\theta(x) \triangleq \frac{x^{\delta-1}(1- x^\delta/\theta^\delta)^{K-2}}{\left(C_\kappa(x,1)\right)^K},
\end{equation*}
we have $\theta'>\theta \Rightarrow f_{\theta'}(x) > f_{\theta}(x), \;\forall x $, 
and $f_\theta(x)$ converges to
$
f(x)\triangleq {x^{\delta-1}}/{\left(C_\kappa(x,1)\right)^K}
$
as $\theta\to\infty$.
Therefore,
\begin{align*}
	\lim_{\theta\to\infty} \theta^\delta \mathsf{P}^\textnormal{c}_{K,1}
		&= \lim_{\theta\to\infty} (K-1)\delta \int_0^\theta f_\theta (x) \d x			\\
		&\pleq{a} (K-1)\delta  \lim_{\theta\to\infty} \int_0^\infty f_\theta (x) \d x		\\
		&\peq{b} (K-1)\delta  \int_0^\infty f (x) \d x,
\end{align*}
where (b) is due to the monotone convergence theorem.
Further, since $f_\theta(x) \leq f(x)$ and $\lim_{x\to \infty} f(x) = 0$,
we have $\lim_{\theta\to\infty}\int_{\theta}^\infty f_\theta (x) \d x = 0$.
This allows replacing the inequality (a) with equality and completes the proof.
\end{IEEEproof}

Proposition~\ref{prop:PcK1theta->Inf} shows, just like in the high-reliability
regime, that $\mathsf{P}^\textnormal{c}_{K,1} = \Theta (\theta^\delta)$
is not affected by the particular choice of $K$ and $\kappa$.
Since $\delta = 2/\alpha$, the coverage probability decays faster when $\alpha$ is
smaller in the high spectral efficiency regime, consistent with intuition.

\section{Intra-cell Diversity (ICD) \label{sec:ICD}}

ICIC creates additional load to the neighboring cells by
reserving the RBs at the coordinated BSs.
The extra load improves the coverage probability as it
reduces the inter-cell interference.

In contrast, with selection combining (SC),
the serving BS transmits to the typical user at $M$ RBs simultaneously,
and the user is covered if the maximum SIR (over the $M$ RBs) exceeds $\theta$.
Like ICIC, SC can also improve the network coverage
at the cost of introducing extra load to the BSs.
Different from ICIC,
SC takes advantage of the intra-cell diversity (ICD)
by reserving RBs at the serving cell.

This section provides a baseline analysis on the coverage with ICD (but without ICIC).

\subsection{General Coverage Expression}

\begin{theorem}
The joint success probability of transmission over $M$ RBs
(without ICIC) is
\begin{equation*}
\mathsf{P}^{\cap \textnormal{c}}_{1,M} = \P(\bigcap_{m=1}^M {\bf S}_{1,m})  
		= \frac{1}{C_1(\theta,M)}.
\end{equation*}
\label{thm:jointsuccess1}
\end{theorem}

Since the proof of Theorem~\ref{thm:jointsuccess1} is
a degenerate version of that of a more general
result stated in Theorem~\ref{thm:jointsuccess},
we defer the discussion of the proof to Section~\ref{sec:ICIC&ICD}.
A similar result was obtained in \cite{ganti:globecom3:2012}
where a slightly different framework was used and
the shadowing effect not explicitly modeled.

Due to the inclusion-exclusion principle,
we have the coverage probability with selection combining over $M$ RBs:

\begin{corollary}[$M$-RB selection combining]
The coverage probability over $M$ RBs
without BS-coordination is
\begin{equation*}
	\mathsf{P}^{\cup \textnormal{c}}_{1,M} = \sum_{m=1}^M (-1)^{m+1} {M\choose m}  \mathsf{P}^{\cap \textnormal{c}}_{1,m},
\end{equation*}
where $\mathsf{P}^{\cap \textnormal{c}}_{1,m}$ is given by Theorem~\ref{thm:jointsuccess1}.
\label{cor:Puc1M}
\end{corollary}

\begin{figure}[t]
\centering
\psfrag{Pc}[c][t]{$\mathsf{P}^{\cup \textnormal{c}}_{1,M}$}
\begin{overpic}[width=\linewidth]{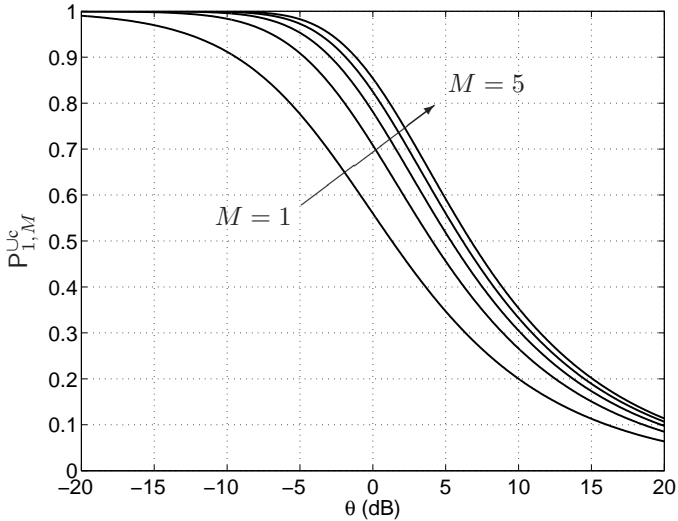}
	\put(43,48){\vector(4,3){20}}
	\put(30,45){$M=1$}
	\put(65,65){$M=5$}
\end{overpic}
\caption{The coverage probability with selection combining over
$M$ RBs without ICIC for $M=1,2,3,4,5$ (lower to upper). Here, $\alpha=4$.}
\label{fig:ICD_M1to5}
\end{figure}

Fig.~\ref{fig:ICD_M1to5} compares the coverage probability
under $M$-RB selection combining, $\mathsf{P}^{\cup \textnormal{c}}_{1,M}$
for $M=1,\cdots,5$.
As expected, the more RBs assigned to the users, the higher
the coverage probability.
Also, similar to the ICIC case, the marginal gain in coverage probability due to ICD
diminishes with $M$.

However, comparing Figs.~\ref{fig:ICD_M1to5}~and~\ref{fig:KcoordCoverage},
we can already observe dramatic difference:
with the same overhead, the coverage gain of ICD looks more evident
than that of ICIC in the high-reliability regime, \ie when $\theta \to 0$.
This observation will be formalized in the following subsection.

\subsection{ICD in the High-Reliability Regime}

\begin{proposition}
Let $\mathsf{P}^{\cap \textnormal{o}}_{1,M} = 1-\mathsf{P}^{\cup \textnormal{c}}_{1,M}$
be the outage probability of the typical user under $M$-RB selection combining.
We have
\begin{equation*}
	\mathsf{P}^{\cap \textnormal{o}}_{1,M}\sim a_M \theta^{M}, \text{ as } \theta\to 0,
\end{equation*}
where $a_M=\frac{\partial^M}{\partial x^M}\left({{_1 F_1} (-\delta;1-\delta; x)}\right)^{-1}\!\big|_{x=0}$ and
${_1 F_1}(a;b;z)$ is the confluent hypergeometric function of
the first kind.
\label{prop:Puo1Mtheta->0}
\end{proposition}

The proof of Proposition~\ref{prop:Puo1Mtheta->0}
can be found in Appendix~\ref{app:Puo1Mtheta->0}.

\begin{remark}
Although Proposition~\ref{prop:Puo1Mtheta->0}
provides a neat expression for the constant in front of $\theta^M$
in the expansion of the outage probability,
numerically evaluating the $M$-th derivative of the reciprocal
of confluent hypergeometric function may not be straightforward.
A relatively simple approach is to resort to Fa\`{a} di Bruno's formula.
Alternatively, one can directly consider \eqref{equ:aMbBM} 
and simplify it by introducing the Bell
polynomial \cite{Johnson02thecurious}: 
\[
a_M =  \sum_{i=1}^M (-1)^i i! \,\textnormal{Bell}_{M,i}(\bar\tau(1),\bar\tau(2),\cdots,\bar\tau(M-i+1)),
\]
where
\begin{equation*}
	\bar\tau(j) \triangleq j! \tau(j) = \frac{(-\delta)_j}{(1-\delta)_j}, 
\end{equation*}
and
\begin{multline*}
\textnormal{Bell}_{m,i}(x_1,\cdots,x_{m-i+1}) = \\ \frac{1}{i!}
\sum_{j_i\geq 1}^{\sum_{i=1}^k j_i=m} {m \choose j_1,\cdots,j_i} x_{j_1}\cdots x_{j_i},
\end{multline*}
which can be efficiently evaluated numerically\footnote{There was a typo in the version published in the December issue of \emph{IEEE Transactions on Wireless Communication} where $\bar\tau$ was mistaken as $\tau$. The typo this corrected in this manuscript.}.
\end{remark}

To better understand Proposition~\ref{prop:Puo1Mtheta->0}
we introduce the following definition of the diversity gain in interference-limited networks,
which is consistent with the diversity gain defined in \cite{net:Haenggi14twc}
and is analogous to the conventional diversity defined (only) for interference-less cases,
see \eg \cite{tse2005fundamentals}. 

\begin{definition}[Diversity (order) gain in interference-limited networks]
The diversity (order) gain, or simply diversity, of interference-limited networks
is
\begin{equation*}
		d \triangleq \lim_{\theta\to 0} \frac{\log\P(\textnormal{SIR}<\theta)}{\log\theta}.
\end{equation*}

\end{definition}

Clearly, Proposition~\ref{prop:Puo1Mtheta->0} shows that a diversity gain can be
obtained by selection combining---in sharp contrast with the results
presented in \cite{net:Haenggi14twc},
where the authors show that there is no such gain in retransmission.
The reason of the difference lies in the different association assumptions.
\cite{net:Haenggi14twc} considers the case where the desired transmitter
is at a fixed distance to the receiver which is independent from the interferer
distribution.
However, this paper assumes that the user is associated with the strongest
BS (on average). In other words, the signal strength from the desired transmitter
and the interference are correlated.
Proposition~\ref{prop:Puo1Mtheta->0} together with \cite{net:Haenggi14twc}
demonstrates that this correlation
is critical in terms of the time/spectral diversity.

\begin{figure}[t]
\centering
\psfrag{Po}[c][t]{$\mathsf{P}^{\cap \textnormal{o}}_{1,M}$}
\begin{overpic}[width=\linewidth]{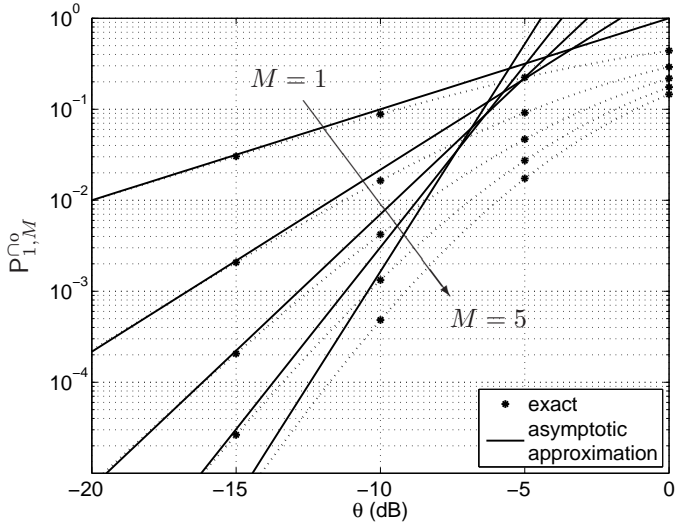}
	\put(43,64){\vector(3,-4){22}}
	\put(35,66){$M=1$}
	\put(65,30){$M=5$}
\end{overpic}
\caption{Asymptotic behavior (and approximation) of the outage probability $\mathsf{P}^{\cap \textnormal{o}}_{1,M}$ with $M$-RB joint transmission for $M = 1,2,3,4,5$ (upper to lower). Here, $\alpha=4$.}
\label{fig:ICD_M1to5asymp}
\end{figure}

Propositions~\ref{prop:PoK1theta->0}~and~\ref{prop:Puo1Mtheta->0}
quantitatively explain the visual contrast between
Figs.~\ref{fig:KcoordCoverage}~and~\ref{fig:ICD_M1to5}
in the high-reliability regime ($\theta\to 0$).
While ICIC reduces the interference by muting nearby interferers,
the number of coordinated BSs only affects the outage probability
by the coefficient and does not change the fact that $\mathsf{P}^\textnormal{o}_{K,1}=\Theta(\theta)$
as $\theta\to 0$.
In contrast, ICD affects the outage probability by both the coefficient and the exponent.

Fig.~\ref{fig:ICD_M1to5asymp}
compares the asymptotic approximation, \ie $a_M \theta^M$,
with the exact expression provided in Corollary~\ref{cor:Puc1M}.
A reasonably accurate match can be found for small $\theta$,
and the range where the approximation is accurate is larger when $M$ is smaller.
Thus, despite the fact that the main purpose of Proposition~\ref{prop:Puo1Mtheta->0}
was to indicate the qualitative behavior of ICD,
the analytical tractability of $a_M$ also provides useful approximations
in applications with small coding rate,
\eg spread spectrum/UWB communication, node discovery, \emph{etc}.

\subsection{ICD in High Spectral Efficiency Regime}

For completeness, we also consider the high spectral efficiency regime where $\theta \to \infty$.

\begin{proposition}
The coverage probability of the typical user under $M$-RB selection combining
satisfies 
\begin{equation}
	\mathsf{P}^\textnormal{c}_{1,M} \sim b_M \theta^{-\delta}, \text{ as } \theta\to \infty,
	\label{equ:Pc1MSimHSE}
\end{equation}
where $b_M = \sum_{m=1}^M (-1)^{m+1} {M \choose m} \frac{\Gamma(m)}{\Gamma(1-\delta)\Gamma(m+\delta)} $.
\label{prop:Pc1Mtheta->Inf}
\end{proposition}

\begin{IEEEproof}
We proceed by (first) considering $\theta^\delta \mathsf{P}^{\cap \textnormal{c}}_{1,M}$.
By Theorem~\ref{thm:jointsuccess1}, we have
\begin{align*}
	\theta^\delta \mathsf{P}^{\cap \textnormal{c}}_{1,M} &= \frac{\theta^\delta}{_2 F_1(m,-\delta;1-\delta;-\theta)}		\\
		&\peq{a} \left( \frac{\theta}{1+\theta}\right )^\delta \frac{1}{_2 F_1(-\delta,1-\delta-m;1-\delta;\frac{\theta}{1+\theta})}	
\end{align*}
where (a) comes from \cite[eqn. 9.131]{GradShteynRyzhik}.
Since $_2 F_1(-\delta,1-\delta-m;1-\delta;1) = {\Gamma(1-\delta)\Gamma(m+\delta)}/{\Gamma(m)}$,
we have $\lim_{\theta\to\infty} \theta^\delta \mathsf{P}^{\cap \textnormal{c}}_{1,M} = \frac{\Gamma(m)}{\Gamma(1-\delta)\Gamma(m+\delta)}$,
which leads to the proposition thanks to Corollary~\ref{cor:Puc1M}.
\end{IEEEproof}

\begin{figure}[t]
\centering
\psfrag{bK and bM}[c][c]{$b_K$ and $b_M$}
\psfrag{bK}[c][c]{$b_K$}
\psfrag{bM}[c][c]{$b_M$}
\psfrag{a}[c][c]{$\alpha$}
\begin{overpic}[width=\linewidth]{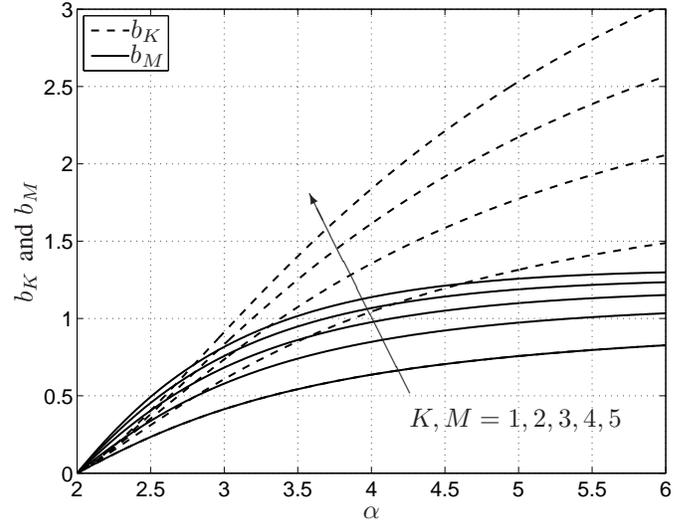}
	\put(60,15){$K,M=1,2,3,4,5$}
	\put(60,20){\vector(-1,2){15}}
\end{overpic}
\caption{$b_K$ ($K=1,2,3,4,5$) in Proposition~\ref{prop:PcK1theta->Inf}
and $b_M$ ($M=1,2,3,4,5$) in Proposition~\ref{prop:Pc1Mtheta->Inf} for different values of $\alpha$,
Here, $\kappa= 1$.}
\label{fig:bKbM_kappa1}
\end{figure}

Comparing Propositions~\ref{prop:PcK1theta->Inf}~and~\ref{prop:Pc1Mtheta->Inf},
we see that unlike the high-reliability regime, the coverage probabilities of ICIC and ICD do \emph{not} have order difference
in the high spectral efficiency regime.
However, the difference in coefficients ($b_K$ and $b_M$) can
also incur significant difference in the coverage probability.
Fig.~\ref{fig:bKbM_kappa1} compares the coefficients for different path loss
exponent $\alpha$ assuming $\kappa = 1$.
Note that $\kappa = 1$ corresponds to the smallest possible $b_K$.
Yet, even so, $b_K$ still dominates $b_M$ for most realistic $\alpha$.
This implies that ICIC is often more effective than ICD in the high spectral efficiency
regime.

\section{ICIC and ICD\label{sec:ICIC&ICD}}

Sections~\ref{sec:ICIC}~and~\ref{sec:ICD} provided
the coverage analysis in cellular networks with ICIC and ICD
separately.
This section considers the scenario where the network takes
advantage of ICIC and ICD at the same time.
In particular, we will evaluate the coverage probability $\mathsf{P}^{\cup \textnormal{c}}_{K,M}$
when the typical user is assigned with $M$ RBs with independent fading
at the serving BS and all the $M$ RBs are also reserved at the
$K-1$ strongest non-serving BSs. 

\subsection{The General Coverage Expression}

In order to derive the coverage probability,
we first generalize Lemma~\ref{lem:xikTildIkLaplace}
beyond Rayleigh fading. In particular, for a generic
fading random variable $H$, we introduce the following definition.

\begin{definition}
For a PLPS $\Xi = \{\xi_i\}$,
let ${I}_k^{H}$ be the interference from the BSs weaker (without fading) than
the $k$-th strongest BS, \ie ${I}_k^{H} = \sum_{i>k} H_i \xi_i^{-1}$,
where $H_i\stackrel{\textnormal{\scriptsize d}}{=} H,\;\forall i\in\mathbb{N},$ are iid.
\end{definition}

Similarly, we define ${I}_\rho^{H} = \sum_{\xi\in\Xi\cap(\rho,\infty)} H_\xi \xi_\rho^{-1}$
to be the interference from BSs with average (over fading) received power less than $\rho^{-1}$.
Then, we obtain a more general version of Lemma~\ref{lem:xikTildIkLaplace} as follows.

\begin{lemma}
For general fading random variables $H\geq 0$ and $\E[H^\delta]<\infty$,
the Laplace transform of ${\xi_k I^H_k}$ is
\begin{equation}
	\L_{\xi_k I^H_k}(s) = \frac{1}{\left(1-\frac{1}{\kappa}+\frac{1}{\kappa}\E_H\left[e^{-sH}+s^\delta H^\delta \gamma(1-\delta,sH)\right]\right)^k}.
	\label{equ:xikTildIkLaplaceGeneral}
\end{equation}
\label{lem:xikTildIkLaplaceGeneral}
\end{lemma}

\begin{IEEEproof}
The proof of the lemma follows exactly that of Lemmas~\ref{lem:LtildeIrho}~and~\ref{lem:xikTildIkLaplace}.
The only difference is that we do not factor in the distribution of the fading random
variable $H$.
More precisely, we can first show
\begin{multline*}
	\L_{\xi_\rho I^H_\rho}(s) = \\
	\exp\left(-\frac{\lambda}{\kappa}\pi\E[S^\delta]\rho^\delta\E_H[e^{-sH}+s^\delta H^\delta \gamma(1-\delta,sH)-1]\right).
\end{multline*}
Then, integrating $\rho$ over the distribution of $\xi_k$ gives the desired result.
\end{IEEEproof}
Note that the condition $\E[H^\delta]<\infty$ in Lemma~\ref{lem:xikTildIkLaplaceGeneral} is sufficient (but not necessary)
to guarantee the existence of the Laplace transform.

As will become clear shortly, for the purpose of this section,
the most important case of $H$ is when $H$ is a gamma random variable with pdf
$f_H(x) = \frac{1}{\Gamma(m)}x^{m-1} e^{-x}$,
where $m\in \mathbb{N}$.
For this case, we have the following lemma.

\begin{lemma}
For $m\in\mathbb{N}$, if $H$ is a gamma random variable with pdf
$f_H(x) = \frac{1}{\Gamma(m)}x^{m-1} e^{-x}$,
\begin{equation*}
\L_{\xi_k I^H_k}(s) = \frac{1}{\left(C_\kappa(s,m)\right)^k}.
\end{equation*}
\label{lem:LxikIkm}
\end{lemma}

Almost trivially based on Lemma~\ref{lem:xikTildIkLaplaceGeneral},
Lemma~\ref{lem:LxikIkm} helps to show the following theorem.

\begin{theorem}
For all $M\in\mathbb{N}$ and $K>1$,
the joint coverage probability over $M$-RBs under
$K$-cell coordination is
\begin{equation*}
\mathsf{P}^{\cap \textnormal{c}}_{K,M} = \P(\bigcap_{m=1}^M {\bf S}_{K,m})  = (K-1) \int_0^1 \frac{(1-x^\delta)^{K-2} \delta x^{\delta-1}}{\left(C_\kappa(\theta x,M)\right)^{K}} \d x.
\end{equation*}
\label{thm:jointsuccess}
\end{theorem}

\begin{IEEEproof}
Let $h_i^m$ be the fading coefficient from the $i$-th strongest (on average)
BS at RB $m$ for $m\in[M]$.
By definition, we have
\begin{equation}
	\mathsf{P}^{\cap \textnormal{c}}_{K,M} 
					= 
					\E_{\Xi} \P\left(h_1^m \xi_1^{-1} >\theta \sum_{i>K} \chi_i h_i^m \xi_i^{-1},\; \forall m\in [M]\right)
\label{equ:PTcKdef}
\end{equation}
Due to the conditional independence (given $\Xi$) across $m$,
\eqref{equ:PTcKdef} can be further simplified as
\begin{align}
	\mathsf{P}^{\cap \textnormal{c}}_{K,M} 
	&= \E_\Xi \prod_{m=1}^M \P\left(h_1^m>\theta\xi_1\sum_{i>K} \chi_i h_i^m \xi_i^{-1}\right) \notag\\
	&= \E_\Xi \E \prod_{m=1}^M \exp\left(-\theta\xi_1\sum_{i>K} \chi_i h_i^m \xi_i^{-1}\right),  \label{equ:E_Xi E} \\
	&= \E_\Xi \E \exp\left(-\theta\xi_1\sum_{i>K} \chi_i \underbrace{\sum_{m=1}^M h_i^m}_{H_i} \xi_i^{-1}\right), \notag
\end{align}
where the inner expectation in \eqref{equ:E_Xi E} is taken over $h_i^m$
for $m\in[M]$ and $i\in\mathbb{N}$,
and due to the independence (across $m$ and $i$) and (exponential) distribution of $h_i^m$,
$H_i$ are iid gamma distributed
with pdf  $f(x)=\frac{1}{\Gamma(M)}x^{M-1} e^{-x}$.

Further, writing $\xi_1$ as $\frac{\xi_1}{\xi_K} \xi_K$
and letting $\Xi_K = \{\xi_i\}_{i=K+1}^{\infty}$,
we obtain the following expression by taking advantage of the statistical independence
shown in Lemma~\ref{lem:PLPSisPPP}:
\begin{equation*}
	\mathsf{P}^{\cap \textnormal{c}}_{K,M} = \E_{\frac{\xi_1}{\xi_K}} \L_{\xi_k I^H_K} \left(\theta \frac{\xi_1}{\xi_K}\right),
\end{equation*}
where $\L_{\xi_K I^H_K}(\cdot)$ is given in Lemma~\ref{lem:LxikIkm}.
The proof is completed by plugging in Lemma~\ref{lem:xi1xik}.
\end{IEEEproof}

Note that although Theorem~\ref{thm:jointsuccess} does not
explicitly address the case $K=1$,
the same proof technique applies to this (easier) case,
where the treatment of the random variable $\xi_1/\xi_K$ is unnecessary since
it has a degenerate distribution ($\equiv 1$).
Thus, the proof of Theorem~\ref{thm:jointsuccess1} is evident and
omitted from the paper.

Due to the inclusion-exclusion principle,
we immediately obtain the following corollary.

\begin{corollary}[$K$-BS coordination and $M$-RB selection combining]
The coverage probability over $M$ RBs
with $K$ BS-coordination is
\begin{equation}
	\mathsf{P}^{\cup \textnormal{c}}_{K,M}  = \sum_{m=1}^M (-1)^{m+1} {M\choose m}  \mathsf{P}^{\cap \textnormal{c}}_{K,m},
\end{equation}
where $\mathsf{P}^{\cap \textnormal{c}}_{K,m}$ is given by Theorem~\ref{thm:jointsuccess}.
\end{corollary}

\begin{figure}[t]
\centering
\psfrag{1-Pc}[c][t]{$\mathsf{P}^{\cap \textnormal{o}}_{K,M}$}
\begin{overpic}[width=\linewidth]{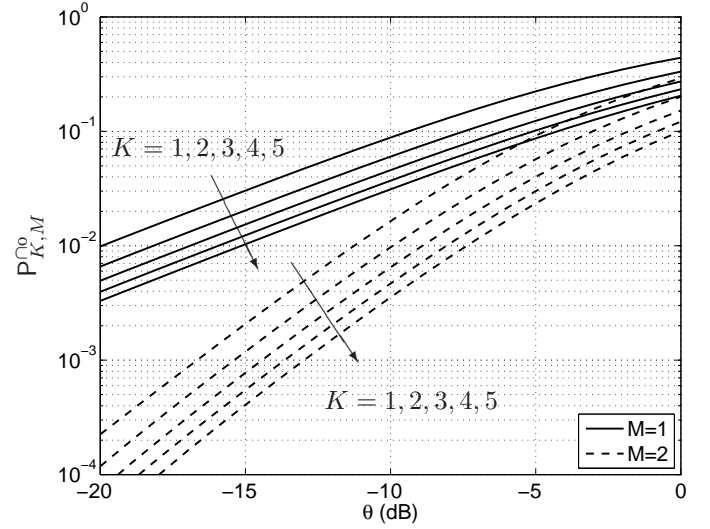}
	\put(28,53){\vector(1,-2){7}}
	\put(13,56){$K=1,2,3,4,5$}
	\put(40,40){\vector(2,-3){10}}
	\put(45,18){$K=1,2,3,4,5$}
\end{overpic}
\caption{The outage probability $\mathsf{P}^{\cap \textnormal{o}}_{K,M}$ under $K$-BS coordination over $M$ RBs
for $K=1,2,3,4,5$ (upper to lower) and $M=1,2$. Here, $\kappa= 1$.}
\label{fig:T1T2}
\end{figure}

\subsection{The High-Reliability Regime}

\begin{proposition}
Let $\mathsf{P}^{\cap \textnormal{o}}_{K,M} = 1-\mathsf{P}^{\cup \textnormal{c}}_{K,M}$
be the outage probability of the typical user under $M$-RB selection combining and $K$-BS coordination.
We have
\begin{equation*}
	\mathsf{P}^{\cap \textnormal{o}}_{K,M}\sim a(K,M) \theta^{M}, \text{ as } \theta\to 0,
\end{equation*}
where $a(K,M)>0\; ,\forall K,M\in\mathbb{N}$.
\label{prop:PuoKMtheta->0}
\end{proposition}

Proposition~\ref{prop:PuoKMtheta->0} combines Proposition~\ref{prop:PoK1theta->0}~and~\ref{prop:Puo1Mtheta->0}.
Its proof is
analogous to that of Proposition~\ref{prop:Puo1Mtheta->0}
(but more tedious) and is thus omitted from the paper.
Proposition~\ref{prop:PuoKMtheta->0} gives quantitative evidence
on why a pure ICD scheme maximizes the coverage probability in
the high-reliability regime.

In Fig.~\ref{fig:T1T2}, we plot the outage probability for difference
number of coordinated cell $K=1,2,3,4,5$ and in-cell diversity $M=1,2$
assuming $\kappa=1$ and observe the consistency with Proposition~\ref{prop:PuoKMtheta->0}.


\section{Numerical Validation\label{sec:numer}}

\subsection{The Effective Load Model\label{subsec:effectiveLoad}}

In Section~\ref{sec:sysPLPSPF}, we introduced the effective load $\kappa$
and modeled the impact of the out-of-cluster coordination on the interference
by independent thinning of the interferer field
with retaining probability $1/\kappa$.
Although, remarkably, $\kappa$ disappears when considering
the diversity order of the network, it is still of interest to
evaluate the accuracy of such modeling in the non-asymptotic regime.
To this end, we set up the following ICIC simulation to validate the effective load model.

We consider the users are distributed as a homogeneous PPP $\Phi_\textnormal{u}$ with density $\lambda_\textnormal{u}$
independent from the BS process $\Phi$.
We assume a single channel and a random scheduling policy
where we pick every user exactly once at a random order.
A picked user is scheduled iff its strongest BSs are not (already) serving another
user or coordinating (\ie being muted) with user(s) in other cell(s)
and its second to $K$-th strongest BSs are not transmitting (serving other users).
Thus, after the scheduling phase, there are at most $\Phi(B)$ users scheduled where
$B\subset \R^2$ is the simulation region since there are at most $\Phi(B)$ serving BSs.
In reality, the number of scheduled users is often much less than $\Phi(B)$ since
1) there is always a positive probability that there are empty cells due to the randomness
in BS and user locations\footnote{This also implies that the full-load assumption
does not hold (exactly) in the simulation.};
2) when $K>1$ some BSs are muted due to coordination.
The ratio between the number of BSs $\Phi(B)$ and
the number of scheduled users (which equals the number of serving BSs)
is consistent with the definition of the effective load $\kappa$
and thus is a natural estimate.
Under lognormal shadowing with standard deviation $\sigma$, we empirically
measured $\kappa$ as in Table~\ref{tab:kappa}.
It is observed that our simulation results in the estimates $\hat\kappa$
that can be well approximated by an affine function of $K$
and the function depends
on the shadowing variance.
The fact that more severe shadowing results in smaller $\kappa$
can be explained in the case $K=1$.
In this case, the only reason that $\kappa>1$ is the existence of empty cells
and the larger $\kappa$ is the more empty cells there are.
Independent shadowing reduces the spatial correlation
of the sizes of nearby Poisson Voronoi cells and thus naturally reduces the variance
of the number of users in each cell, resulting in a smaller number of empty cells.

\begin{table}[htbp]
  \centering
  \caption{Estimated $\kappa$}
    \begin{tabular}{cccccc}
    \toprule
    $K$     & 1     & 2     & 3     & 4     & 5 \\
    \midrule
    $\sigma=0$~dB   & 1.0101 & 1.7166 & 2.3640 & 2.9889 & 3.6018 \\
    $\sigma=6$~dB  & 1.0022 & 1.6385 & 2.1904 & 2.7145 & 3.2206 \\
    $\sigma=10$~dB  & 1.0008 & 1.6129 & 2.1096 & 2.5730 & 3.0152 \\
    \bottomrule
    \end{tabular}%
  \label{tab:kappa}%
\end{table}%

\begin{figure*}[t]
	\begin{minipage}[t]{.47\linewidth}
		\centering
		\psfrag{Pc}[c][t]{$\mathsf{P}^\textnormal{c}_{K,1}$}
		\begin{overpic}[width=\linewidth]{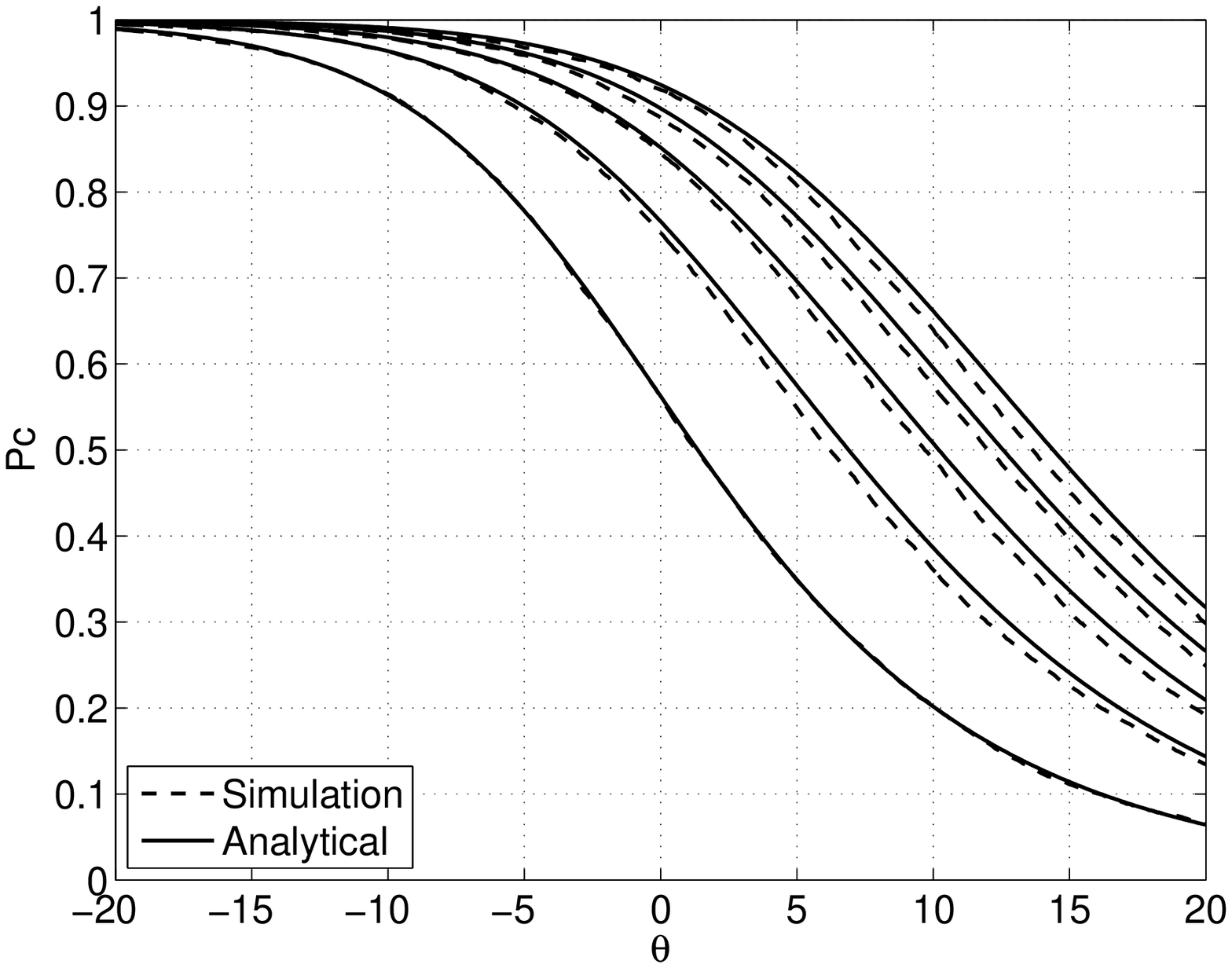}
			\put(15,45){$K=1,2,3,4,5$}
			\put(50,47){\vector(4,1){30}}
		\end{overpic}
		\subcaption{$\sigma=0$~dB}\label{fig:Kcoord_Sim}
	\end{minipage}
	\begin{minipage}[t]{.47\linewidth}
		\centering
		\psfrag{Pc}[c][t]{$\mathsf{P}^\textnormal{c}_{K,1}$}
		\begin{overpic}[width=\linewidth]{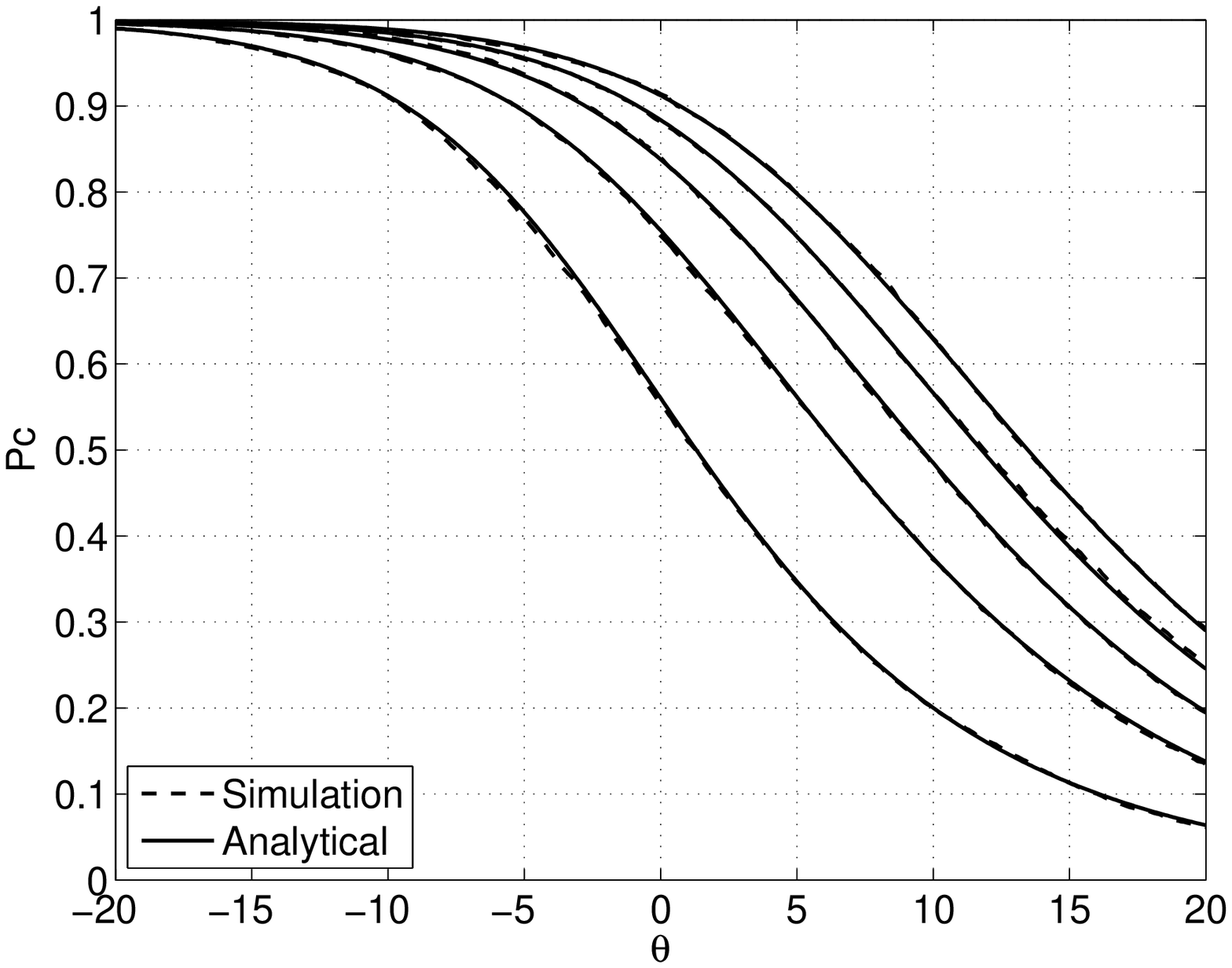}
			\put(15,45){$K=1,2,3,4,5$}
			\put(50,47){\vector(4,1){30}}
		\end{overpic}
		\subcaption{$\sigma=10$~dB}\label{fig:Kcoord_Sim_sigmav10}
	\end{minipage}
		\caption{The coverage probability comparison between the analytical coverage probability
		derived in Theorem~\ref{thm:KCoord}
		and the simulation results with $K=1,2,3,4,5$.
		Here, the BS density $\lambda =1$, user density $\lambda_\textnormal{u} = 10$.
		Lognormal shadowing with variance $\sigma^2$ (in dB$^2$) is considered.}
\end{figure*}

Fig.~\ref{fig:Kcoord_Sim} compares the coverage probability under $K$-BS coordination predicted by
Theorem~\ref{thm:KCoord} using the estimated $\kappa$ from Table~\ref{tab:kappa} with the simulation results.
We picked the case where $\sigma = 0$~dB since this is the \emph{worst case}
in terms of matching analytical results with the simulation
due to the size correlation of Poisson Voronoi cells.
To see this more clearly, consider the case $K=1$, 
where the simulation and analysis match almost completely in the figure.
The match is expected but not entirely trivial since the process of transmitting BSs
is no longer a PPP.
More specifically, a BS is transmitting iff there is at least one user in its Voronoi cell,
\ie the ground process $\Phi$ is thinned by the user process $\Phi_\textnormal{u}$.
However, the thinning events are spatially dependent due to the dependence in
the sizes of the Voronoi cells.
As a result, clustered BSs are less likely to be serving users at the same time and thus
the resulting transmitting BS process is more \emph{regular} than a PPP.
Yet, Fig.~\ref{fig:Kcoord_Sim} shows that the deviation from a PPP is small when
$\lambda_\textnormal{u}=10\lambda$.\footnote{In fact,
the match for $K=1$ is still quite good for smaller user densities, say $\lambda_\textnormal{u} = 5\lambda$.} 

Shadowing breaks the spatial dependence of
the interfering field (transmitting BSs) and consequently improves the accuracy of the analysis.
When $\sigma=10$~dB, the difference between the simulated coverage probability and
the one predicted in Theorem~\ref{thm:KCoord} are almost visually indistinguishable (Fig.~\ref{fig:Kcoord_Sim_sigmav10}).
These results validate the effective load model for analyzing ICIC in the non-asymptotic regime.

%
%

\subsection{ICIC-ICD Trade-off}

The analyses in Sections~\ref{sec:ICIC},~\ref{sec:ICD}~and~\ref{sec:ICIC&ICD}
shows the significantly different behavior of ICIC and ICD schemes despite the fact
that both the schemes improves the coverage probability through generating extra load
in the system.
In particular, Propositions~\ref{prop:PoK1theta->0},~\ref{prop:Puo1Mtheta->0}~and~\ref{prop:PuoKMtheta->0}
show that when $\theta\to 0$, ICD will have a larger impact on the coverage probability due
to the diversity gain.
In contrast, Propositions~\ref{prop:PcK1theta->Inf}~and~\ref{prop:Pc1Mtheta->Inf} suggest that when $\theta\to\infty$, 
ICIC will be more effective since the linear gain is typically larger (Fig.~\ref{fig:bKbM_kappa1}).
Intuitively, a ICIC-ICD combined scheme should present a trade-off between the performance
in these two regimes.

To make a fair comparison between different ICIC-ICD combined schemes,
we need to control their load on the system in terms of RBs used.
By the construction of the model, we observe that the load introduced by ICIC is the effective load $\kappa$
times the load without ICIC
since $1/\kappa$ is the fraction of active transmitting BSs, which, in the single-channel case,
is proportional to the number (or, density) of users being served.
Similarly, the load introduced by ICD is $M$ times the load without ICD
since $M$ RBs are grouped to serve a single user (while without ICD they could be used to serve
$M$ users). 
Thus, under both $K$-BS coordination and $M$-RB selection combining,
the system load is proportional to $\kappa M$
which we term \emph{ICIC-ICD load factor}.

\begin{figure}[t]
\centering
\psfrag{Pc}[c][t]{$\mathsf{P}^{\cup \textnormal{c}}_{K,M}$}
\includegraphics[width=\linewidth]{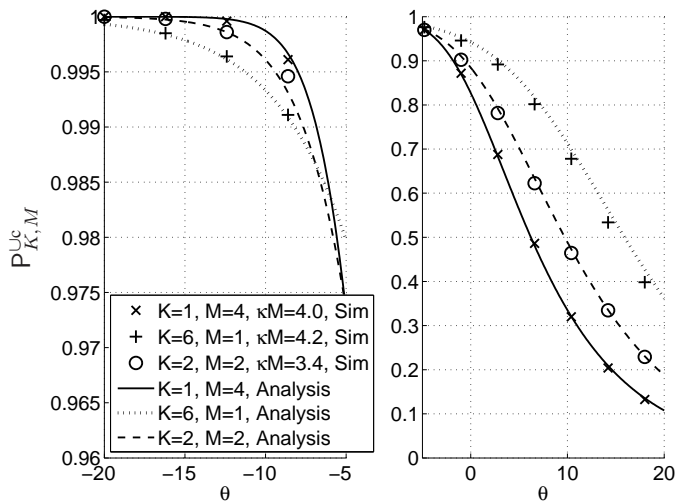}
\caption{The coverage probability under $K$-BS coordination
and $M$-RB selection combining $\mathsf{P}^{\cup \textnormal{c}}_{K,M}$ with
different combinations $(K,M)$.
Here, $\alpha=4$, $\sigma=0$~dB. 
The left figure shows the part for $\theta\in[-20~\textnormal{dB},-5~\textnormal{dB}]$
and the right figure for $\theta\in[-5~\textnormal{dB},20~\textnormal{dB}]$.
}
\label{fig:ICICICDSimAna_dbl}
\end{figure}

Fig.~\ref{fig:ICICICDSimAna_dbl}
plots the coverage probability of three ICIC-ICD combined schemes
with different but similar ICIC-ICD load factor $\kappa M$ using both the analytical result and simulation.
As is shown in the figure, a hybrid ICIC-ICD scheme (\ie with $K,M>1$)
provides a trade-off between the good performance of ICIC and ICD in the two asymptotic regimes.
In general, a hybrid scheme could provide the highest coverage probability
for intermediate $\theta$, and the crossing point depends on all the system parameters.


%

\subsection{ICIC-ICD-Load Trade-off}

Another more fundamental trade-off is between the load and the ICIC-ICD combined schemes.
In other words, how to find the optimal combination $(K,M)$ that takes the load into account.
While the complexity of this problem prohibits a detailed exploration
in this paper, we give a simple example to explain the trade-off.

Assume all the users in the network are transmitting at the same rate $\log(1+\theta)$
and the network employs the random scheduling procedure as described in Section~\ref{subsec:effectiveLoad}.
Then the (average) throughput of the typical scheduled user is $\log(1+\theta) \mathsf{P}^{\cup \textnormal{c}}_{K,M}$ in the interference-limited network.
Under the ICIC and ICD schemes,
the number of user being served per RB is (on average) $1/\kappa M$ times those who can be served in the
baseline case without ICIC and ICD.
Therefore, for fixed $\theta$,
the spatially averaged (per user) throughput is proportional to $\mathsf{P}^{\cup \textnormal{c}}_{K,M}/\kappa M$.
Intuitively, it is the product of the probability of a random chosen user being scheduled ($\propto 1/\kappa M$)
and the probability of successful transmission ($\mathsf{P}^{\cup \textnormal{c}}_{K,M}$).
Then we can find optimal combination 
\begin{equation}
(K^*,M^*) = \argmax_{(K,M)\in\mathbb{N}^2} \frac{\mathsf{P}^{\cup \textnormal{c}}_{K,M}}{\kappa M}
\label{equ:KstarMstar}
\end{equation}
using exhaustive search.
Alternatively, we can enforce an outage constraint and find the optimal $(K,M)$ combination such
that 
\begin{equation}
(K^*,M^*) = \argmax_{(K,M)\in \mathbb{N}^2} \frac{\mathsf{P}^{\cup \textnormal{c}}_{K,M}}{\kappa M} \sf{1}_{[1-\epsilon,1]}(\mathsf{P}^{\cup \textnormal{c}}_{K,M})
\label{equ:KstarMstarOutage}
\end{equation}

\begin{figure}[t]
\centering
\psfrag{Pc}[c][t]{$\mathsf{P}^{\cup \textnormal{c}}_{K,M}$}
\includegraphics[width=\linewidth]{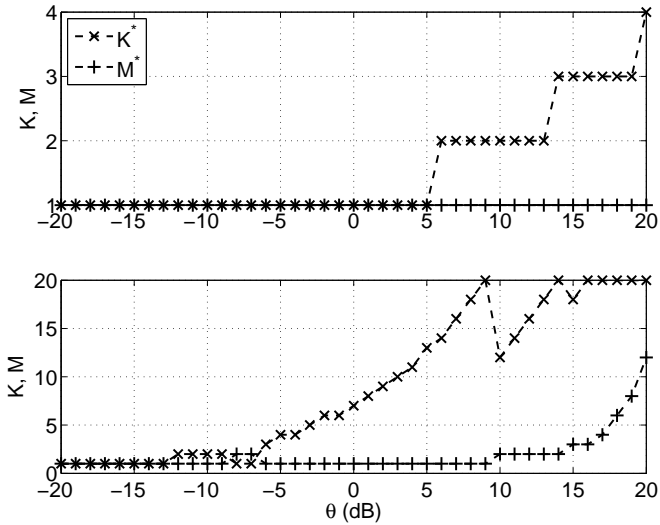}
\caption{
The optimal $(K^*,M^*)$ as a function of $\theta$.
The top subfigure is optimized for average throughput, see \eqref{equ:KstarMstar}.
The bottom subfigure is optimized for average throughput under an outage constraint, see \eqref{equ:KstarMstarOutage}
with $\epsilon = 0.05$.
$10$~dB shadowing is considered.
}
\label{fig:KstarMstar}
\end{figure}

Fig.~\ref{fig:KstarMstar} plots the exhaustive search result for $(K^*,M^*)$
defined in \eqref{equ:KstarMstar} and \eqref{equ:KstarMstarOutage}.
In the simulation, we limit our search space for both $K$ and $M$ to $\{1,2,\cdots,20\}$
and we use the affine function $\kappa = \eta_0 + \eta_1 K$ to
approximate $\kappa$, which turns out to be an accurate fit in our simulation
(see the data in Table~\ref{tab:kappa}).

Fig.~\ref{fig:KstarMstar} shows that as $\theta$ increases,
it is beneficial to increase $K$.
This is consistent with the result derived in 
Propositions~\ref{prop:PcK1theta->Inf}~and~\ref{prop:Pc1Mtheta->Inf}
and Fig.~\ref{fig:bKbM_kappa1}, which show
that ICIC is more effective in improving coverage probability for large $\theta$.
If there is no outage constraint,
it is more desirable to keep both $K$ and $M$ (and thus the load factor) small.
This is true especially for small $\theta$ since the impact of $(K,M)$ on
$\mathsf{P}^{\cup \textnormal{c}}_{K,M}$ is small ($\mathsf{P}^{\cup \textnormal{c}}_{K,M}\approx 1$)
but both $\kappa$ (a function of $K$) and $M$ linearly affect the load factor and thus the average throughput.

The incentive to increase $(K,M)$ is higher if an outage constraint is imposed.
Although it is still more desirable to increase $K$ (both due to its usefulness in the high-spectral efficiency
regime and its smaller impact on the load factor), the increase in $M$ also has a non-trivial impact:
a slight increase in $M$ could significantly reduce the optimal value of $K$.
This is an observation of practical importance,
since the cost of increasing $K$ is usually much higher than that of increasing $M$
due to the signaling overhead that ICIC requires.

\section{Conclusions \label{sec:conclu}}

This paper provides explicit expressions for the coverage probability
of inter-cell interference coordination (ICIC) and intra-cell diversity (ICD)
in cellular networks modeled by a homogeneous Poisson point process (PPP).
Examining the high-reliability regime,
we demonstrate a drastically different behavior of ICIC and ICD
despite their similarity in creating extra load in the network.
In particular, ICD, under the form of selection combining (SC),
provides diversity gain
while ICIC can only linearly affect the outage probability
in the high-reliability regime.
In contrast, in the high-spectral efficiency regime,
ICIC provides higher coverage probability for realistic path loss exponents.
All the analytical results derived in the paper are invariant to the
network density and the shadowing distribution.

The fact that ICD under selection combining provides diversity gain
in cellular networks even with temporal/spectral interference correlation
contrasts with the corresponding results in \emph{ad hoc} networks,
where \cite{net:Haenggi14twc} shows no such gain exists.
This shows that the spatial dependence between the desired transmitter
and the interferers is critical in harnessing the diversity gain.

In the non-asymptotic regime,
we propose an effective load model to analyze the effect of ICIC.
The model is validated with simulations and proven to be very accurate.
Using these analytical results,
we explore the fundamental trade-off between ICIC-ICD and system load
in cellular systems.

\appendices

\section{Proof of Proposition~\ref{prop:PoK1theta->0} \label{app:PoK1theta->0}}

\begin{IEEEproof}
We prove the theorem by calculating $\lim_{\theta\to 0}\frac{\mathsf{P}^\textnormal{o}_{K,1}}{\theta}$.
First, consider the case where $K=1$.
Since $\mathsf{P}^\textnormal{o}_{K,1} = 1-\mathsf{P}^\textnormal{c}_{K,1}$, we have
\begin{equation*}
	\lim_{\theta\to 0}\frac{\mathsf{P}^\textnormal{o}_{K,1}}{\theta}
	 = \lim_{\theta\to 0} \frac{C_\kappa(\theta,1)-1}{\theta C_\kappa(\theta,1)}
	 = \lim_{\theta\to 0} C'_\kappa(\theta,1),
\end{equation*}
where $C'_\kappa(x,1) = \frac{\d}{\d x}C_\kappa(x,1)$,
and the last equality is due to L'Hospital's rule and the fact that $C_1(0,1) = 1$.
Moreover, we have
\begin{equation}
\lim_{\theta\to 0} C'_\kappa(\theta,1) = \frac{1}{\kappa} C'_1(0,1)= \frac{1}{\kappa} \frac{\delta}{1-\delta}
\label{equ:C'101}
\end{equation}
due to 
the series expansion of the Gauss hypergeometric function ${_2 F_1} (a,b;c;z) = \sum_{n=0}^{\infty}\frac{(a)_n (b)_n}{(c)_n} \frac{z^n}{n!}$.
Thus, we proved \eqref{equ:PoK1Sim} is true for $K=1$.

For $K\geq 2$, by Theorem~\ref{thm:KCoord}, we have
\begin{multline*}
	\mathsf{P}^\textnormal{o}_{K,1} = \\
			(K-1)\delta\int_0^1 \left(1-\frac{1}{(C_\kappa(\theta x,1))^K}\right)
								(1-x^\delta)^{K-2}x^{\delta-1}\d x,
\end{multline*}
where the integral, by change of variable $y = \theta x$, can be written as
\begin{equation*}
	\frac{1}{\theta^\delta}\int_0^\theta \Delta_K(y)
								\left(1-\frac{y^\delta}{\theta^\delta}\right)^{K-2}y^{\delta-1}\d y,
\end{equation*}
where $\Delta_K(y) = 1-(C_\kappa(y,1))^{-K}$.
Therefore, we have
\begin{equation*}
	\lim_{\theta\to 0}\frac{\mathsf{P}^\textnormal{o}_{K,1}}{\theta}
			=	\lim_{\theta\to 0} \frac{(K-1)\delta}{\theta^{\delta+1}}\int_0^\theta \Delta_K(y)
								\left(1-\frac{y^\delta}{\theta^\delta}\right)^{K-2}y^{\delta-1}\d y,
\end{equation*}
where the RHS can be simplified by (repetitively) applying L'Hospital's rule as follows:
\begin{align}
&\phantom{={}} \lim_{\theta\to 0} \frac{(K-1)\delta}{\theta^{\delta+1}}\int_0^\theta \Delta_K(y)
		\left(1-\frac{y^\delta}{\theta^\delta}\right)^{K-2}y^{\delta-1}\d y		\notag\\
&= \lim_{\theta\to 0} \frac{(K-2)_2 \delta^2}{(\delta+1)\theta^{2\delta+1}} \int_0^\theta \Delta_K(y)
		\left(1-\frac{y^\delta}{\theta^\delta}\right)^{K-3}y^{2\delta-1}\d y		\notag\\
&= \cdots	\notag\\
&= \frac{(K-1)! \delta^{K-1}}{\prod_{k=1}^{K-2}(k\delta+1)} \lim_{\theta\to 0} \frac{1}{\theta^{(K-1)\delta+1}}
		\int_0^\theta \Delta_K(y) y^{(K-1)\delta -1}\d y		\notag\\
&= \frac{(K-1)! \delta^{K-1}}{\prod_{k=1}^{K-1}(k\delta+1)} \lim_{\theta\to 0} \frac{\Delta_K(\theta)}{\theta},
\label{equ:limPoK1/theta_ll}
\end{align}
where
\begin{equation*}
	\lim_{\theta\to 0} \frac{\Delta_K(\theta)}{\theta} = \lim_{\theta\to 0} \frac{(C_\kappa(\theta,1))^K-1}{\theta (C_\kappa(\theta,1))^K}.
\end{equation*}
Note that $\lim_{\theta\to 0} C_\kappa(\theta,1) = 1$ and thus $\lim_{\theta\to 0} \frac{\Delta_K(\theta)}{\theta} = \lim_{\theta\to 0} \frac{(C_\kappa(\theta,1))^K-1}{\theta }$, which by L'Hospital's rule can be further simplified
as $K\lim_{\theta\to 0} (C_\kappa(\theta,1))^{K-1} C'_\kappa(\theta,1)$.
Therefore,
thanks to \eqref{equ:C'101}, we have
\begin{equation}
	\lim_{\theta\to 0} \frac{\Delta_K(\theta)}{\theta} = K C'_\kappa(0,1) = \frac{K}{\kappa} \frac{\delta}{1-\delta}.  \label{equ:limDeltaK/theta}
\end{equation}

Combining \eqref{equ:limPoK1/theta_ll} and \eqref{equ:limDeltaK/theta} completes the proof.
\end{IEEEproof}

\section{Proof of Proposition~\ref{prop:Puo1Mtheta->0} \label{app:Puo1Mtheta->0}}

In order the prove Proposition~\ref{prop:Puo1Mtheta->0},
we first introduce two useful lemmas.
Letting $\mathsf{D}^k_u = \frac{\partial^k}{\partial u^k}$,
the following lemma states a simple algebraic fact which will turn out
to be useful in the asymptotic analysis.

\begin{lemma}\label{lem:Dkuc1uM}
For any $c\in\R$,
we have
\begin{multline*}
	\mathsf{D}^k_u\left.\left(1-\frac{1}{c(1+u)}\right)^M\right|_{u=0} = \\
		\sum_{j=1}^k {M \choose j}{k \choose j} j! (M)_{k-j} (-1)^{k-j} \left(1-\frac{1}{c}\right)^{M-j}.	
\end{multline*}
\end{lemma}
\begin{IEEEproof}
First, expressing $\left(1-\frac{1}{c(1+u)}\right)^M$
as $(cu+c-1)^M(\frac{1}{c(1+u)})^M$, by the Leibniz rule,
we can expand the $k$-th order derivative as
\begin{multline*}
\mathsf{D}^k_u\left(1-\frac{1}{c(1+u)}\right)^M = \\
		\sum_{j=0}^k {k \choose j} \mathsf{D}^j_u(cu+c-1)^M \mathsf{D}^{k-j}_u\frac{1}{c^M (1+u)^M},
\end{multline*}
where
\begin{align*}
\left. \mathsf{D}^j_u(cu+c-1)^M\right|_{u=0} 
			&= \mathsf{D}^j_u\left. \sum_{m=1}^M {M \choose m} (cu)^m (c-1)^{M-m} \right|_{u=0}	\\
			&= {M \choose j} j! c^j (c-1)^{M-j},
\end{align*}
and
\begin{equation*}
	\mathsf{D}^{k-j}_u\left.\frac{1}{c^M (1+u)^M} \right|_{u=0} = \frac{(-1)^{k-j}}{c^M} (M)_{k-j}.
\end{equation*}
This gives the desired expansion.
\end{IEEEproof}

Thanks to Lemma~\ref{lem:Dkuc1uM},
we have the following result.

\begin{lemma}
Given $n$ arbitrary nonnegative integers $k_1,k_2,\cdots,k_n\in\mathbb{N}\cup\{0\}$ and $A_n\triangleq\sum^n_{i=1} k_i \geq 1$,
we have
\begin{equation*}
	\sum_{m=1}^{M} {M \choose m} (-1)^{m+A_n} \prod_{i=1}^n (m)_{k_i} =
	\begin{cases}
	0,						&\text{if } A_n < M		\\
	M!,						&\text{if } A_n = M
	\end{cases}
\end{equation*}
for all $M\in\mathbb{N}$.
\label{lem:sumMmmki}
\end{lemma}

\begin{IEEEproof}
We prove the lemma by induction.
First, consider the case where $n=1$.
Then for all $k_1>0$, we have
\begin{align*}
	\hspace{2em}&\hspace{-2em}\sum_{m=1}^M (-1)^{k_1+m} {M\choose m} (m)_{k_1} 		\\
					&= \sum_{m=1}^M (-1)^{m} {M\choose m} \mathsf{D}^{k_1}_u\left.\frac{1}{(1+u)^m}\right|_{u=0}	\\
					&= \mathsf{D}^{k_1}_u \left.\left(	\left(1-\frac{1}{1+u}\right)^M -1	\right)\right|_{u=0}		\\
					&= \mathsf{D}^{k_1}_u \left.\frac{u^M}{(1+u)^M}\right|_{u=0},
\end{align*}
where the $k_1$-th derivative can be expanded by the Leibniz rule, \ie
\begin{equation*}
	\mathsf{D}^{k_1}_u\frac{u^M}{(1+u)^M} = \sum_{i=0}^{k_1} {k_1\choose i} 
								\left(\mathsf{D}^i_u u^M \right) 
								\left(\mathsf{D}^{k_1-i}_u \frac{1}{(1+u)^M} \right),
\end{equation*}
which is $0$ when $u=0$ if $k_1<M$.
When $k_1=M$, the only non-zero term in
the sum is the one with $i=k_1=M$ and thus is $\mathsf{D}^{k_1}_u\frac{u^M}{(1+u)^M}|_{u=0}=M!$.
Therefore, the lemma is true for $n=1$.

Second, we prove it for the case $0<A_n < M$
with general $n$ by induction.
Assume
\begin{multline*}
\sum_{m=1}^{M} {M \choose m} (-1)^{m+A_{n-1}} \prod_{i=1}^{n-1} (m)_{k_i} = \\
\sum_{m=1}^{M} {M \choose m} (-1)^{m}  \prod_{i=1}^{n-1} \left.\mathsf{D}^{k_i}_{u_i} \left(\frac{1}{1+u_i}\right)^m \right|_{\substack{u_i=0 \\ i\in[n]}} = 	0,		
\end{multline*}
for all $n-1$ nonnegative integers $\{k_i\}_{i=1}^{n-1}$ with $0<A_{n-1}=\sum_{i=1}^{n-1} k_i < M$.
Then, we consider the case for $n$,
move all the $\mathsf{D}^{k_i}_{u_i}$ to the front and,
analogous to the $n=1$ case, obtain
\begin{multline}
\sum_{m=1}^{M} {M \choose m} (-1)^{m+\sum\limits^{n}_{i=1} k_i} \prod_{i=1}^{n} (m)_{k_i} = \\
\left.\left(\prod_{i=1}^{n} \mathsf{D}^{k_i}_{u_i}\right) \left(1-\frac{1}{\prod_{i=1}^{n}(1+u_i)}\right)^M
	\right|_{\substack{u_i=0 \\ i\in[n]}}.
\label{equ:nderi}
\end{multline}
Expanding only the $k_n$-th order derivative using Lemma~\ref{lem:Dkuc1uM},
we can express \eqref{equ:nderi} as
\begin{equation}
\left(\prod_{i=1}^{n-1} \mathsf{D}^{k_i}_{u_i}\right)
\sum_{j=1}^{k_n} a_{k_n,j,M} \left(1-\frac{1}{\prod_{i=1}^{n-1}(1+u_i)}\right)^{M-j},\label{equ:nderiexpand}
\end{equation}
where $a_{k_n,j,M}={M \choose j}{k_n \choose j} j! (M)_{k_n-j} (-1)^{k_n-j}$
is independent from $k_i$ and $u_i$ for all $i\in[n-1]$.
We then can move the derivative operators inside the summation.
Further, since $j\leq k_n < M-\sum_{i=1}^{n-1} k_i$, we have
$\sum_{i=1}^{n-1} k_i < M-j$ for all $j$ in the summation, which
leads to the observation that
\[
\left.
\left(\prod_{i=1}^{n-1} \mathsf{D}^{k_i}_{u_i}\right)
\left(1-\frac{1}{\prod_{i=1}^{n-1}(1+u_i)}\right)^{M-j}\right|_{\substack{u_i=0 \\ i\in[n]}} =0,\; \forall j\in[k_n]
\]
by our assumption on the $n-1$ case.
Thus the lemma is proved for the case $0<A_n < M$
for all $n\in\mathbb{N}$.

For the case $A_n = M$,
we see, by the first part of the proof, that 
\[\left.
\left(\prod_{i=1}^{n-1} \mathsf{D}^{k_i}_{u_i}\right)
\left(1-\frac{1}{\prod_{i=1}^{n-1}(1+u_i)}\right)^{M-j}\right|_{\substack{u_i=0 \\ i\in[n]}}
\]
can be non-zero only if $j=k_n$.
Thus \eqref{equ:nderiexpand} can be simplified
to 
\[\left.
{M \choose k_n} k_n ! \left(\prod_{i=1}^{n-1} \mathsf{D}^{k_i}_{u_i}\right) \left(1-\frac{1}{\prod_{i=1}^{n-1}(1+u_i)}\right)^{M-k_n}\right|_{\substack{u_i=0 \\ i\in[n]}},
\]
which is $M!$ if we assume the lemma is true for $n-1$.
Since $n$ is arbitrarily chosen, the lemma is proved for
all $n\in\mathbb{N}$.
\end{IEEEproof}

With Lemmas~\ref{lem:Dkuc1uM}~and~\ref{lem:sumMmmki},
we are able to proceed with the proof of Proposition~\ref{prop:Puo1Mtheta->0}
as follows.

\begin{IEEEproof}[Proof (of Proposition~\ref{prop:Puo1Mtheta->0})]
By Corollary~\ref{cor:Puc1M}, we have
\begin{align}
\mathsf{P}^{\cap \textnormal{o}}_{1,M} 
		&= 1-\sum_{m=1}^M (-1)^{m+1} {M\choose m} \frac{1}{C_1(\theta,m)}	\notag\\
		&= \sum_{m=1}^M (-1)^{m+1} {M\choose m} \left(1-\frac{1}{C_1(\theta,m)}\right).		\label{equ:Puo1m1stblood}
\end{align}
We then proceed the proof by considering the Taylor expansion
of ${1}/{C_1(x,m)}$ at $x = 0$.
To find the $n$-th derivative of ${1}/{C_1(x,m)}$
we treat $\frac{1}{C_1(x,m)}$ as a composite of $f(x)=x^{-1}$
and $C_1(x,m)$, where
the derivatives of $C_1(x,m)$ is available by the series expansion
of hypergeometric function mentioned before.
Then, by Fa\`{a} di Bruno's formula \cite{Johnson02thecurious}, we have
\begin{multline}
 \left.\mathsf{D}^n_x\left(\frac{1}{C_1(x,m)}\right)\right|_{x=0}=	\\
		\sum_{{\bf b}\in{\cal B}_n}\frac{n!  (\sum_{i=1}^{n} b_i)!}{\prod_{i=1}^n (b_i!)} \prod_{i=1}^n\left( \frac{(m)_i (-\delta)_i}{(1-\delta)_i i!} \right)^{b_i},
\label{equ:1/cderi}
\end{multline} 
where ${\cal B}_n$ is the set of $n$-tuples of non-negative
integers $(b_i)_{i=1}^n$ with $\sum_{i=1}^n{i b_i}=n$,
and ${\bf b} = (b_i)_{i=1}^n$.
\eqref{equ:1/cderi} directly leads to the Taylor expansion
of ${1}/{C_1(\theta,m)}$, which combined with \eqref{equ:Puo1m1stblood}
leads to a series expansion of $\mathsf{P}^{\cap \textnormal{o}}_{1,M}$
as function of $\theta$,
\begin{multline*}
\mathsf{P}^{\cap \textnormal{o}}_{1,M} 
=\\
\sum_{m=1}^M (-1)^{m} {M\choose m}
\sum_{n=1}^\infty \theta^n
\sum_{{\bf b}\in{\cal B}_n}\frac{ (\sum\limits_{i=1}^{n} b_i)!}{\prod\limits_{i=1}^n (b_i!)} \prod_{i=1}^n\big( (m)_i\tau(i) \big)^{b_i},
\end{multline*}
where $\tau(i) \triangleq \frac{ (-\delta)_i}{(1-\delta)_i i!}$.
Rearranging the sums and products in expression above yields
$\mathsf{P}^{\cap \textnormal{o}}_{1,M} = \sum_{n=1}^\infty a_n \theta^n$,
where
\begin{equation*}
	a_n = \sum_{{\bf b}\in{\cal B}_n}\frac{ (\sum\limits_{i=1}^{n} b_i)!}{\prod\limits_{i=1}^n (b_i!)} \prod_{i=1}^n \big(\tau(i)\big)^{b_i}
\sum_{m=1}^M (-1)^{m} {M\choose m}
\prod_{i=1}^n \big((m)_i\big)^{b_i}.
\end{equation*}

Recall that ${\bf b}\in{\cal B}_n$ indicates $\sum_{i=1}^n{i b_i}=n$.
By Lemma~\ref{lem:sumMmmki}, we have $a_n = 0$ for all $n<M$,
\ie $\mathsf{P}^{\cap \textnormal{o}}_{1,M} = O(\theta^M)$ as $\theta \to 0$.
Further, Lemma~\ref{lem:sumMmmki} helps us to obtain
the coefficient in front of $\theta^M$,
\ie
\begin{equation}
		a_M = \sum_{{\bf b}\in{\cal B}_M}\frac{ M! (-1)^{\sum_{i=1}^{M} b_i} (\sum_{i=1}^{M} b_i)!}{\prod_{i=1}^M (b_i!)} \prod_{i=1}^M \big(\tau(i)\big)^{b_i},
\label{equ:aMbBM}
\end{equation}
which leads to the concise expression in the proposition
by reusing Fa\`{a} di Bruno's formula.
\end{IEEEproof}

\bibliographystyle{IEEEtran}

\end{document}